\documentclass[a4paper,11pt]{article}
\pdfoutput=1 % if your are submitting a pdflatex (i.e. if you have
\usepackage{jheparxiv} % for details on the use of the package, please
\usepackage[T1]{fontenc} % if needed

\usepackage{twistor}
\usepackage{tensor}
\usepackage{amsmath}
\usepackage{amssymb}
\usepackage{mathrsfs}
\usepackage{bm}
\usepackage{graphicx}

\newcommand{\eps}{\epsilon}
\newcommand{\veps}{\varepsilon}

\newcommand{\psum}{\sideset{}{'}\sum}
\newcommand{\pprod}{\sideset{}{'}\prod}
\newcommand{\g}{\mathfrak{g}}
\newcommand{\mfk}[1]{\mathfrak{#1}}
\newcommand{\msf}[1]{\mathsf{#1}}
\newcommand{\Res}[1]{\underset{#1}{\text{Res}}\,}
\newcommand{\cK}{\mathcal{K}}
\newcommand{\mS}{\mathcal{S}}
\newcommand{\jt}{\tilde{\jmath}}
\newcommand{\cS}{\mathcal{S}}

\title{Celestial double copy from the worldsheet}

\author[a]{Eduardo Casali}
\author[b]{and Atul Sharma}

\affiliation[a]{Center for Quantum Mathematics and Physics (QMAP)
and\\Department of Physics, University of California,
Davis, CA 95616 USA}
\affiliation[b]{The Mathematical Institute\\
University of Oxford, Woodstock Road, OX2 6GG, United Kingdom}

\emailAdd{ecasali@ucdavis.edu}
\emailAdd{atul.sharma@maths.ox.ac.uk}

\abstract{Using the ambitwistor string, we compute tree-level celestial amplitudes for biadjoint scalars, Yang-Mills and gravity to all multiplicities. They are presented in compact CHY-like formulas with operator-valued scattering equations and numerators acting on a generalized hypergeometric function. With these we extend the celestial double copy to tree-level amplitudes with arbitrary number of external states. We also show how color-kinematics duality is implemented in celestial amplitudes and its interpretation in terms of a generalized twisted cohomology theory.}

\begin{document} 
\maketitle
\flushbottom

\section{Introduction}
\label{sec:intro}

The perturbative S-matrix of quantum field theories can have many interesting properties and relations. If we take these relations to reflect some fundamental property of the respective QFTs, we expect them to survive in some form after a change of basis in the Hilbert space. The S-matrix is usually computed in a basis of plane waves for external particles due to its simplicity, but other bases might be more useful in answering certain questions. In~\cite{Cheung:2016iub,Pasterski:2016qvg,Pasterski:2017kqt,Pasterski:2017ylz} it was argued that a basis of external particles transforming as conformal primaries on the sphere at the null boundary of Minkowski space is the most appropriate one to study a potential holographic CFT living on this sphere. Amplitudes with conformal wavefunctions as external states were dubbed \textit{celestial amplitudes}. These can be obtained from amplitudes computed in the usual plane wave basis by a Mellin transform. This has the effect of mixing the UV and the IR. Yet universal properties of amplitudes, such as soft limits and IR factorization~\cite{Cheung:2016iub,Donnay:2018neh,Fan:2019emx,Pate:2019mfs,Adamo:2019ipt,Puhm:2019zbl,Guevara:2019ypd,Law:2019glh,Fotopoulos:2019vac,Banerjee:2020kaa,Fan:2020xjj,Nandan:2019jas,Gonzalez:2020tpi}, survive this change of basis albeit in a different guise. Recently, it was shown in~\cite{Casali:2020vuy} that relations between the S-matrices of Yang-Mills and gravity known as the double copy also survives this change of basis, at least up to four point amplitudes. 

While an all-multiplicity proof could be found using the methods employed in~\cite{Casali:2020vuy}, which relied on position space Feynman rules, they soon become cumbersome. From our experience with the tree-level amplitudes in the usual plane wave basis we know that there should be more compact descriptions of the S-matrix. For certain classes of celestial amplitudes in four dimensions all-multiplicity formulas already exist in the literature: tree-level Yang-Mills MHV and NMHV amplitudes were computed in~\cite{Schreiber:2017jsr}. These computations are done by performing Mellin transforms on the energies of the external particles which, while doable for some low point amplitudes including loop~\cite{Albayrak:2020saa,Gonzalez:2020tpi} and string~\cite{Stieberger:2018edy} amplitudes, become highly involved for generic amplitudes. CFT inspired methods have also been used to holographically constrain MHV amplitudes~\cite{Banerjee:2020zlg,Banerjee:2020vnt} and it might be that further understanding of the celestial OPEs~\cite{Pate:2019lpp,Banerjee:2020kaa,Fotopoulos:2019tpe,Ebert:2020nqf} will lead to better methods.

% some one loop amplitudes~\cite{} and.... Looking at the simplest example of MHV amplitudes already shows how complicated these amplitudes can be. The results in in \cite{schreiber2018tree} involve  generalized hypergeometric functions on Gr$(4,n)$ and Gelfand-A hypergeometric functions. While usual tree-level amplitudes are rational functions of the Mandesltam and polarizations. We expect this difficulty to only grow as one moves away from MHV amplitudes and even more for gravitational amplitudes at higher multiplicities.
% On the other hand the CHY formulas~\cite{} give a compact representation for the whole tree-level S-matrix of Yang-Mills and gravity using the scattering equations~\cite{}. In~\cite a Mellin transform of the CHY formula was used to study celestial soft theorems, but the Mellin integrations were not carried out explicitly.

Here we choose to use a true and tested method to generate compact expressions for massless S-matrices, the ambitwistor string~\cite{Mason:2013sva}. This is a worldsheet theory from which the CHY formulas~\cite{Cachazo:2013hca,Cachazo:2013iea} can be obtained as correlation functions of vertex operators in a chiral CFT. These are a generalization of the original Berkovits-Witten twistor string~\cite{Witten:2003nn,Berkovits:2004jj}. Being string theories, the sum over Feynman diagrams is replaced by an integral over the moduli space of punctured spheres providing a very compact representation of the tree-level S-matrix. A Mellin transform of the CHY formula was used in~\cite{Adamo:2019ipt} to study celestial soft theorems, but the Mellin integrations were not carried out explicitly. Here we carry those out explicitly and check it against a first principles derivation of it using the ambitwistor string with vertex operators in the conformal basis. This gives a framework for celestial massless S-matrices making use of the \textit{celestial} scattering equations in analogy with the usual CHY formula. The amplitudes are written as operators acting on the Mellin transform of a scalar contact vertex which we carry out explicitly in terms of generalized hypergeometric functions.   

Another important feature of the CHY formulas and the ambitwistor string is that the double copy relation between Yang-Mills and gravity amplitudes is made manifest. The double copy is a procedure to obtain gravitational amplitudes from a ``square'' of Yang-Mills amplitudes~\cite{Bern:2010ue}. While in the ambitwistor string the double copy amounts to a substitution rule, using traditional methods it relies on kinematical numerators satisfying the so-called color-kinematics duality~\cite{Bern:2008qj}, that is, they obey relations analogous to Jacobi identities among color factors. Using the formulas computed using the ambitwistor string for celestial amplitudes we show that tree-level celestial Yang-Mills and gravity amplitudes are related by a double copy, generalizing the procedure given for low points in~\cite{Casali:2020vuy}. Moreover, by interpreting the numerators given by the ambitwistor string as cohomology classes in a generalization of twisted cohomology, we can uplift the results of~\cite{Mizera:2019blq} to the celestial case. In doing this we find a natural generalization of color-kinematics duality to celestial amplitudes and show how to obtain color-kinematical dual celestial numerators from the ambitwistor string.

An extra motivation to study celestial amplitudes is as a toy model for understanding double copy and color-kinematics duality for amplitudes around curved backgrounds. Low multiplicity amplitudes have been computed for plane wave backgrounds in~\cite{Adamo:2017nia,Adamo:2017sze,Adamo:2018mpq} and shown to have a double copy structure, with some higher point results recently given in~\cite{Adamo:2020syc,Adamo:2020qru,Adamo:2020yzi}. But computations 
still relied heavily on the leftover momentum conservation and special properties of plane wave backgrounds. Ambitwistor string computations have also provided new formulas for amplitudes in AdS spacetimes~\cite{Eberhardt:2020ewh,Roehrig:2020kck} which have a structure very close to the one we find for celestial amplitudes. We would like to find a notion of double copy that can be more easily generalized to curved spaces and valid for a larger class of spacetimes including asymptotically AdS spacetimes. 

%Recently, ambitwistor string computations in AdS spacetimes have provided new formulas for amplitudes~\cite{Eberhardt:2020ewh,Roehrig:2020kck} which have a structure very close to the one we find for celestial amplitudes. One of the goals of this paper was also to set up a framework for the double copy using celestial amplitudes which can be more straightforwardly generalized beyond flat space.\enote{redundant with things said in conclusion, could just take this paragraph out.}

\paragraph{Summary of results:}
\begin{itemize}
\item We compute compact, $n$-point formulas for tree-level celestial amplitudes of biadjoint scalars, gluons and gravitons. They are given in a universal form as operator-valued numerators acting on the Mellin transform of the scalar contact vertex. Analogous to the case with external plane waves we can represent the amplitudes either as a sum over trivalent graphs~\eqref{trigraph}, or as integrals over the moduli space of $n$-punctured Riemann surfaces localized to the operator-valued \textit{celestial scattering equations} \eqref{Anmain}. \\
\item As a consequence of the computation given above, we prove that the celestial double copy introduced in~\cite{Casali:2020vuy} is valid for amplitudes at all multiplicities. \\
\item We introduce a generalization of twisted cohomology to operator-valued twisted forms which are the relevant objects in the CHY formulas for celestial amplitudes. We use this, together with a straightforward generalization of the results in~\cite{Mizera:2019blq}, to define color-kinematics duality for celestial amplitudes and show how to obtain color-kinematical dual numerators from the ambitwistor string numerators.
\end{itemize}

This paper is organized as follows: we start in section~\ref{sec:review} by reviewing some facts about celestial amplitudes, CHY formulas, double copy and twisted homology to make this paper more self-contained. The reader familiar with these topics can safely skip these sections. Section~\ref{sec:cse} contains the formulas for celestial amplitudes with explanations about its constituents. Next, in section~\ref{sec:ambitwistor}, we introduce the ambitwistor model and go through the calculation used to obtain the formulas introduced previously. In section~\ref{sec:cel_ck} we introduce a generalization of the twisted cohomology to operator-valued forms, give a definition of color-kinematics duality for celestial amplitudes and show how numerators obtained from the ambitwistor string naturally obey this duality. We finish with some discussion about this framework and possible further generalizations in section~\ref{sec:end}.   

% respectively. However, these are rather unwieldy. Also, we have the CHY formulas, and we could hope that, akin to David's model in \cite{Roehrig:2020kck}, we might find localization in Mellin space or connections to integrable systems. Here, we explicitly work with a theory of massless scalars.

%%%%%%%%%%%%%%%%%%%%%%%%%%%%%%%%%%%%%
%%%%%%%%%%%%%%%%%%%%%%%%%%%%%%%%%%%%%

\section{Review}
\label{sec:review}

In order to make this paper self-contained we quickly review in this section some background on recent technology used in the study of amplitudes. We keep the reviews short and focused on what is needed in the rest of the paper. In subsection~\ref{subsec:Celestial} we review some facts about celestial amplitudes, in subsection~\ref{subsec:CHY} CHY formulas and the scattering equations are introduced. In subsection \ref{subsec:ckreview} we recall the basics of color-kinematics duality, and in subsection~\ref{subsec:t_cohomology} we review how the CHY formulas and double copy are to be interpreted in light of twisted cohomology on the moduli space of punctured Riemann spheres.

\subsection{Celestial amplitudes}
\label{subsec:Celestial}

Celestial amplitudes are obtained by Mellin transforming the usual momentum space amplitudes to make manifest their transformations under the conformal group of the celestial sphere at null infinity $\scri^+$. This is essentially a change of basis~\cite{Pasterski:2017kqt} on the Fock space from the plane wave basis into a basis of conformal primary wavefunctions\footnote{There's evidence that more states than the ones obtained from the Mellin transform are necessary to describe the quantum theory on $\scri$, see~\cite{Pate:2019lpp}.}

Let $X$ be a point in $\R^{1,d+1}$, with $D:=d+2$, and denote the massless scalar plane wave as $\e^{\im s\omega q\cdot X}$ where $\omega$ denotes its energy, $q\in S^d$ is a direction on the celestial sphere, and $s=\pm1$ denotes if the particle is outgoing or incoming. The scalar conformal primary wavefunction is obtained by Mellin transforming the energy $\omega$,
\begin{equation}\label{wavefunction_s}
    \phi_\Delta^s(X;q) = \mathcal{M}(\e^{\im s\omega q\cdot X-\veps\omega})=\int_0^\infty \frac{\d \omega}{\omega}\,\omega^{\Delta}\,\e^{\im s\omega q\cdot X-\veps\omega} = \frac{(-\im\,s)^\Delta\,\Gamma(\Delta)}{(-q\cdot X-\im\,s\,\veps)^\Delta}\,,
\end{equation}
giving an external wavefunction that lives on the celestial sphere with a new quantum number: the conformal dimension $\Delta$. Conformal wavefunctions for external gluons and gravitons also exist but here we will work directly with the Mellin transform of the plane waves,
\begin{equation}\label{wavefunctions_YM_g}
 \begin{array}{cc}
  a^{\msf{a}, s }_{\mu,\Delta}=\msf{T^a}\,\epsilon_\mu\,\phi_\Delta^ s (X;q)\;\;\;\; & h^{ s }_{\mu\nu,\Delta}=\epsilon_{\mu}\,\tilde\eps_\nu\,\phi_\Delta^ s (X;q)\,,
 \end{array}
\end{equation}
which are given in terms of the scalar wavefunctions, the color generators $\msf{T^a}$, and polarizations $\eps^\mu, \tilde\eps^\mu$ that only depend on $q^\mu$~\cite{Pasterski:2017ylz}. These wavefunctions transform as conformal primaries up to gauge transformations~\cite{Pasterski:2017kqt}. Wavefunctions for massive particles have also been worked out~\cite{Law:2020tsg,Narayanan:2020amh} but we won't have anything to say about massive particles. Amplitudes are in principle computed using \eqref{wavefunction_s} and \eqref{wavefunctions_YM_g} as external wavefunctions, but it is more convenient to simply Mellin transform the usual amplitudes computed with plane waves. A celestial amplitude is formally given by
\begin{equation}
 \mathcal{A}(\{\Delta_i, q_i,  s_i \})=\int_{\R_+^n}\prod_{i=1}^n\frac{\d \omega_i}{\omega_i}\,\omega^{\Delta_i}\,A(\{\omega_i,q_i,s_i\})\,.
\end{equation}
Explicitly carrying out the Mellin transforms can quickly become unwieldy as the number of external particles increases. In the following section we'll introduce the CHY representation of amplitudes which gives a very compact formula for massless $n$-point amplitudes. 

%%%%%%%%%%%%%%%%%%%%%%%%%%%%%%%%%%%%%

\subsection{CHY formulas}
\label{subsec:CHY}

The CHY formulas present the $D$-dimensional, tree-level massless S-matrix of several quantum field theories~\cite{Cachazo:2013hca,Cachazo:2013iea,Cachazo:2014xea} as an integral formula. A generic CHY formula is written as
\begin{equation}\label{CHY}
 A_n=\delta^D\left(\sum_{i}^n k_i\right)\int_{\cM_{0,n}}\frac{\d^{n}z}{\text{vol SL}(2,\C)} \;\mathcal{I}(\{g,k,\epsilon,z\})\;\tilde{\mathcal{I}}(\{g,k,\tilde\epsilon,z\})\;\pprod_i\bar\delta(E_i)\,.
\end{equation}
The integral is taken over the moduli space of $n$-punctured Riemann spheres $\mathcal{M}_{0,n}$ which carries an action of the group $\text{SL}(2,\C)$ denoted by the factor of $1/\text{vol SL}(2,\C)$ in the measure. Concretely, this implies we can fix the position of any 3 punctures, for example $\{z_1,z_2,z_3\}$, removing the associated differentials $\d z_i$ and introducing the Jacobian $(z_1-z_2)(z_2-z_3)(z_3-z_1)$ in the integral. The final result is invariant under the choice of which particles are fixed. The numerators $\mathcal{I}$ $(\tilde{\mathcal{I}})$ are rational functions of the external data $\{k,\epsilon,\tilde\eps\}$, couplings $g$ and coordinates $z_i$ on $\mathcal{M}_{0,n}$. Different choices of numerators correspond to different choices of theories. For example, taking
 \begin{equation}\label{PT}
  \mathcal{I}=\tilde{\mathcal{I}}=\text{PT}_n=\frac{\text{Tr}(\msf{T}^{\msf{a}_1}\msf{T}^{\msf{a}_2}\cdots\msf{T}^{\msf{a}_n})}{(z_1-z_2)(z_2-z_3)\cdots(z_{n}-z_1)}+\text{perm.}
 \end{equation}
gives the S-matrix for a cubic biadjoint scalar. Another interesting numerator is the reduced Pfaffian,
 \begin{equation}\label{Pf}
  \text{Pf} '\Psi_n=2\frac{(-1)^{i+j}}{(z_i-z_j)}\text{Pf }(\Psi^{ij}_{ij})\,,
 \end{equation}
with $\text{Pf}$ the Pfaffian of the $2n\times 2n$ matrix,
\begin{equation}
 \Psi=\left(\begin{array}{cc}
             A & -C^T\\
             C & B
            \end{array}
\right)\,,
\end{equation}
with two lines and two columns removed, denoted by $\Psi^{ij}_{ij}$. Its components are given by the matrices
\begin{equation}
\begin{array}{ccc}
 A_{ij}=\begin{cases}
        \dfrac{k_i\cdot k_j}{z_i-z_j} & i\neq j\\
        0 & i=j
        \end{cases}\,,
        &\,
        B_{ij}=\begin{cases}
        \dfrac{\epsilon_i\cdot \epsilon_j}{z_i-z_j} & i\neq j\\
        0 & i=j
        \end{cases}\,,
        &\,
        C_{ij}=\begin{cases}
        \dfrac{\epsilon_i\cdot k_j}{z_i-z_j} & i\neq j\\
        -\sum_{l\neq i}\dfrac{\epsilon_i\cdot k_l}{z_i-z_l} & i=j
        \end{cases}
        \end{array}\,.
\end{equation}
Taking $\mathcal{I}=\text{Pf}'\Psi_n$ and $\tilde{\mathcal{I}}=\text{PT}_n$ as numerators the CHY formula gives the S-matrix for external gluons in Yang-Mills. Using a Pfaffian for both, $\mathcal{I}=\text{Pf}'\Psi(k,\eps)$ and $\tilde{\mathcal{I}} = \text{Pf}'\Psi(k,\tilde\eps)$, gives the S-matrix for gravitational amplitudes in NS-NS gravity.

The universality of the CHY representation is due to the presence of the scattering equations $E_i$~\cite{Cachazo:2013gna}, which are the last ingredient of \eqref{CHY} to be explained. They appear as the arguments of the delta functions in $\prod'_i\bar{\delta}(E_i)$. These delta functions are taken as holomorphic delta functions, that is $\bar\delta(E_i)=\bar\partial(z)^{E_i}$, effectively fixing the contour of integration to the solutions of the scattering equations. The symbol $\prod'$ means that delta functions for three equations should be omitted from the product and another Jacobian of the form $(z_i-z_j)(z_j-z_k)(z_k-z_i)$ should be added. The scattering equations themselves are
\begin{equation}
 E_i=\sum_{j\neq i}\frac{k_i\cdot k_j}{z_i-z_j}\,.
\end{equation}
The delta functions impose them as constraints on the $z_i$'s, completely localizing the integration over $\mathcal{M}_{0,n}$. There are generically $(n-3)!$ points in $\mathcal{M}_{0,n}$ which solve the scattering equations. Denoting the solutions of the scattering equations by $\sigma_i$, the amplitudes above can be written as
\begin{equation}\label{CHY_amp}
 A_n=\delta^D\left(\sum_{i=1}^n k_i\right)\sum_{i=1}^{(n-3)!}\left.\frac{\mathcal{I}\;\tilde{\mathcal{I}}}{\Phi}\right|_{\sigma_i}
\end{equation}
with $\Phi=\text{det }\partial_jE_i$, the Jacobian coming from the delta functions.

In the CHY formulas the sum over trivalent graphs is replaced by the fully localized integral over $\mathcal{M}_{0,n}$, making manifest properties which are hard to see from the Feynman diagrammatic expansion. CHY formulas also exist for many theories beyond the ones reviewed above~\cite{Cachazo:2014xea,Cachazo:2014nsa}; many can be seen as originating from an unconventional string theory, called the ambitwistor string~\cite{Mason:2013sva,Casali:2015vta}. The advantage of having a worldsheet theory like the ambitwistor string is the ease of generalizations to other settings, e.g. higher loops~\cite{Adamo:2013tsa,Geyer:2015bja,Adamo:2015hoa,Geyer:2015jch,Geyer:2016wjx,Geyer:2018xwu} and curved backgrounds~\cite{Adamo:2014wea,Adamo:2017sze,Adamo:2018hzd,Adamo:2018ege,Azevedo:2016zod,Chandia:2015sfa}, which in turns leads to new formulas for amplitudes in these settings. In section~\ref{sec:ambitwistor} we'll show how CHY-like formulas for celestial amplitudes originate from the ambitwistor string.

%%%%%%%%%%%%%%%%%%%%%%%%%%%%%%%%%%%%%

\subsection{Color-kinematics duality and double copy}
\label{subsec:ckreview}

\begin{figure}\centering
 \includegraphics[scale=0.8]{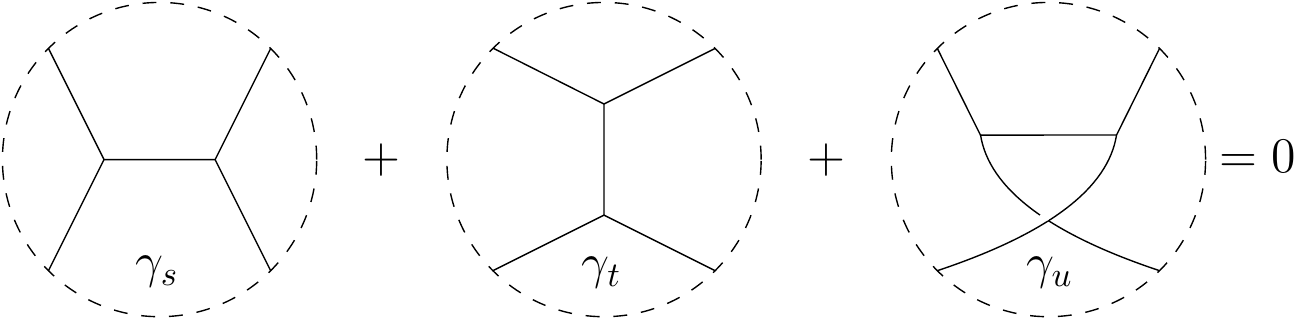}
 \caption{Graphs related by BCJ moves}\label{fig:BCJ_graphs}
\end{figure}

The original double copy~\cite{Bern:2010ue} is a prescription to obtain gravitational amplitudes from suitable squares of Yang-Mills amplitudes. There are now many pairs of theories known to have amplitudes related by a double copy prescription, as well as several proposals for double copy of non-linear solutions. See~\cite{Bern:2019prr} for a recent review of the field. Here we'll present the aspects of the original double copy which we'll need in the rest of the paper. Given a tree-level Yang-Mills amplitude it can be represented as a sum over trivalent graphs by opening up four point interactions in the Feynman diagrams,
\begin{equation}\label{YM_tri}
 A^{\text{YM}}_n=\delta^D\left(\sum_{i=1}^n k_i\right)\sum_{\gamma\in\Gamma}\frac{c_\gamma\,n_\gamma}{\prod_{e\in\gamma}P_e}\,.
\end{equation}
Here $\Gamma$ is the set of trivalent graphs, $c_\gamma$ are color numerators carrying the gauge group information, $n_\gamma$ are kinematical numerators which are polynomials in the external momenta and polarizations, and $P_e$ are the propagators associated to each edge $e$ of the graph $\gamma$.  Color numerators corresponding to graphs which differ only by BCJ moves on some subgraph, see figure~\ref{fig:BCJ_graphs}, obey identities inherited from the usual Jacobi identity, 
\begin{equation}
 c_{\gamma_s}+c_{\gamma_t}+c_{\gamma_u}=0\,.
\end{equation}
If the four point vertices are opened up appropriately then, at tree-level, kinematical numerators can always be found such that they satisfy identities analogous to the ones satisfied by the color numerators. That is they satisfy the Jacobi-like relation,
\begin{equation}
n_{\gamma_s}+n_{\gamma_t}+n_{\gamma_u}=0\,.
\end{equation}
These kind of numerators are called color-kinematical dual numerators~\cite{Bern:2008qj}. Substituting in \eqref{YM_tri} the color numerators by a set of color-kinematical dual numerators $\tilde n_\gamma$ (obtained from $n_\gamma$ by replacing $\eps$ with $\tilde\eps$) yields another amplitude,
\begin{equation}\label{G_tri}
 A^{\text{G}}_n=\delta^D\left(\sum_{i=1}^n k_i\right)\sum_{\gamma\in\Gamma}\frac{n_\gamma\,\tilde n_\gamma}{\prod_{e\in\gamma}P_e}\,,
\end{equation}
which turns out to be a gravitational amplitude. The requirement that numerators obey color-kinematics duality ensures that~\eqref{G_tri} is invariant under linear diffeomorphisms. This is the double copy prescription.

The CHY formulas reviewed in the previous section also have a double copy structure. There, the analogue of color numerators is the Parke-Taylor factor $\text{PT}$ and the analogue of kinematical numerators is the Pfaffian $\text{Pf}'\Psi$. The sum over trivalent graphs is replaced by an integral over the moduli space and double copy becomes a simple substitution rule. Color-kinematics seems to be disconnected from the double copy in this case, but they turn out to be intimately related through twisted cohomology on $\cM_{0,n}$. We review this in the following subsection. 

\subsection{Twisted cohomology}
\label{subsec:t_cohomology}

The ingredients in the CHY formula have an interesting interpretation in terms of a cohomological theory on the moduli space of punctured Riemann spheres $\mathcal{M}_{0,n}$. Here we'll quickly go over the relevant details of the constructions in~\cite{Mizera:2019gea,Mizera:2019blq} that we'll generalize in section~\ref{sec:cel_ck}. Proofs of the statements presented below, as well as details of the computations in the context of amplitudes can be found in~\cite{Mizera:2019gea,Mizera:2019blq,Mizera:2016jhj}. Other mathematical details can be found in the original  mathematical literature~\cite{10.2307/1969099,aomoto2011theory,aomoto1977structure,yoshida2013hypergeometric}. 

We start by defining a meromorphic one-form in $\mathcal{M}_{0,n}$ using the scattering equations:
\begin{equation}\label{eq:conn}
 \omega = \psum_i E_i\,\d z_i\,,\qquad E_i = \sum_{j\neq i}\frac{k_i\cdot k_j}{z_{ij}}\,,
\end{equation}
where the prime in $\sum'$ denotes that three points have been fixed to account for the $\text{SL}(2,\C)$ invariance. Using momentum conservation one can show that the form~\eqref{eq:conn} doesn't depend on which three points were fixed and that it only has simple poles along the boundaries of $\mathcal{M}_{0,n}$. With this in hand we define the \textit{twisted} de Rham operators
\begin{equation}
 \nabla_\pm=\d \pm \omega\,.
\end{equation}
%Both are nilpotent differentials $(\nabla_\pm)^2=0$ acting on $\Omega^\bullet(\cM_{0,n})$, the complex of differential forms on $\mathcal{M}_{0,n}$. We use them to define twisted cohomology groups,
%\begin{equation}
 %H_{\nabla_\pm}^\bullet(\mathcal{M}_{0,n})=\frac{\text{Im }\nabla_\pm}{\text{Ker }\nabla_\pm}\,.
%\end{equation}
%Generically, i.e.\ for generic momenta, the only non-vanishing groups are the middle-dimensional ones $H_{\nabla_\pm}^{n-3}(\mathcal{M}_{0,n})$ with $(n-3)!$ independent generators.
Both are flat connections $[\nabla_\pm,\nabla_\pm]=0$ on certain line bundles $\mathcal{L}$ and its dual $\mathcal{L}^\vee$ over $\mathcal{M}_{0,n}$ called local systems. Sections of $\cL$ $(\cL^\vee)$ are given by functions on $\cM_{0,n}$ that are locally covariantly constant in $\nabla_+$ ($\nabla_-$). The operator $\nabla_+$ acts naturally on $\Omega^\bullet(\mathcal{M}_{0,n},\mathcal{L})$, the complex of differential forms on $\mathcal{M}_{0,n}$ with coefficients in $\mathcal{L}$, and squares to zero $(\nabla_+)^2=0$. We use it to define \textit{twisted} cohomology groups, 
\begin{equation}
H^\bullet(\mathcal{M}_{0,n},\mathcal{L})=\frac{\text{Im }\nabla_+}{\text{Ker }\nabla_+}\,.
\end{equation}
Generically, i.e. for generic momenta, the only non-vanishing group is the middle-dimensional one $H^{n-3}(\mathcal{M}_{0,n},\mathcal{L})$ with $(n-3)!$ independent generators. There is an analogous construction for cohomology groups with values on $\mathcal{L}^\vee$ which we omit.

%Taking $\d\mu$ a top holomorphic form on $\mathcal{M}_{0,n}$, the numerators $\varphi_+=\mathcal{I}\,\d\mu$ and $\varphi_-=\tilde{\mathcal{I}}\,\d\mu$ can be interpreted as elements of the twisted cohomology groups $\varphi_\pm\in H_{\nabla_\pm}^{n-3}(\mathcal{M}_{0,n})$. Moreover, amplitudes in the CHY formalism can be interpreted as a bilinear pairing between these cohomology groups:
%\begin{equation}
 %A_n=\langle \varphi_+|\varphi_-\rangle=\delta^D\left(\sum_{i=1}^n k_i\right)\sum_{i=1}^{(n-3)!}\left.\frac{\mathcal{I}\;\tilde{\mathcal{I}}}{\Phi}\right|_{\sigma_i}\,,
%\end{equation}
%giving the intersection number of $\varphi_+$ and $\varphi_-$.

Taking $\d\mu$ a top holomorphic form on $\mathcal{M}_{0,n}$, the numerators $\varphi_+=\mathcal{I}\,\d\mu$ and $\varphi_-=\tilde{\mathcal{I}}\,\d\mu$ can be interpreted as elements of the twisted cohomology groups $\varphi_+\in H^{n-3}(\mathcal{M}_{0,n},\mathcal{L})$ and $\varphi_-\in H^{n-3}(\mathcal{M}_{0,n},\mathcal{L}^\vee)$. Moreover, amplitudes in the CHY formalism can be interpreted as a bilinear pairing between these cohomology groups:
\begin{equation}
A_n=\langle \varphi_+|\varphi_-\rangle=\delta^D\left(\sum_{i=1}^n k_i\right)\sum_{i=1}^{(n-3)!}\left.\frac{\mathcal{I}\;\tilde{\mathcal{I}}}{\Phi}\right|_{\sigma_i}\,,
\end{equation}
giving the intersection number of $\varphi_+$ and $\varphi_-$.

An alternative evaluation of this pairing can be given where the contours around the scattering equations are deformed, picking up contributions only from the boundaries of $\mathcal{M}_{0,n}$. A codimension 1 boundary is reached when the $n$-punctured sphere degenerates into a nodal surface given by two punctured spheres connected by a node. Higher codimension boundaries are reached with further dengenerations of these spheres into surfaces with more nodes. The deepest boundaries arise when all that is left is a nodal surface composed of spheres with three marked points. These spheres are connected through these marked points in such a way that no closed cycle can be drawn that passes through the nodal points. That is, the nodal surface resembles a tree-graph with only trivalent vertices. In fact, one can label these deepest boundaries by trivalent trees whose punctures are distributed along the external edges. Different boundaries correspond to different assignments of labels for the external punctures. 

The end result is that the contour given by the scattering equations can be deformed to pick contributions from the deepest boundaries of the moduli space. This presents the amplitude as a sum over the boundaries of $\mathcal{M}_{0,n}$ labelled by trivalent graphs,
\begin{equation}
 A_n=\delta^D\left(\sum_{i=1}^n k_i\right)\sum_{\gamma\in\Gamma}\frac{(\varphi_+)_\gamma\, (\varphi_-)_\gamma}{P_\gamma}\,.
\end{equation}
In this expression,
\begin{equation}\label{residue}
 (\varphi_\pm)_\gamma=\Res{v_\gamma}\varphi_{\pm}
\end{equation}
are residues of the top meromorphic forms $\varphi_\pm$ along the boundary divisor $v_\gamma$ labelled by the trivalent graph $\gamma$. This representation is similar to the field theory one~\eqref{YM_tri} and~\eqref{G_tri} but here the numerators are guaranteed to obey color-kinematics. This follows from global residue theorems on $\mathcal{M}_{0,n}$ as explained in~\cite{Mizera:2019blq}. 

\begin{figure}\centering
 \includegraphics{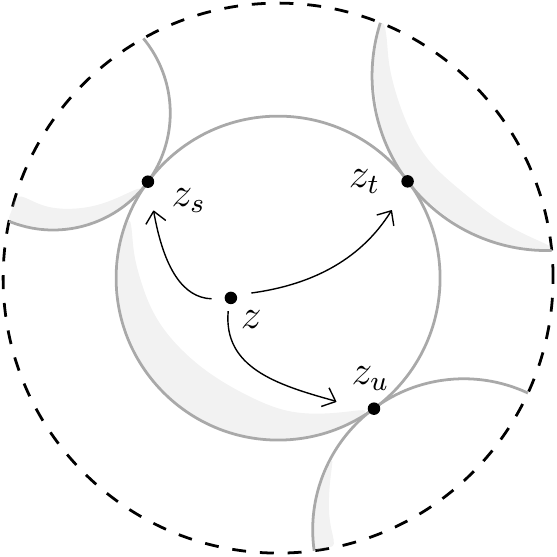}\caption{Neighborhood of the three degenerations related by BCJ moves.}\label{fig:sphere3}
\end{figure}

The general argument can be summarized as follows: take a triple of trivalent graphs $\gamma_s,\gamma_t,\gamma_u$ differing only on a subgraph connecting four edges as shown in figure~\ref{fig:BCJ_graphs}. These three divisors can be seen as arising from the same corner of $\mathcal{M}_{0,n}$ where a four-punctured sphere degenerates as one of its punctures, $z$, approaches one of the other three punctures, $z_s,z_t,z_u$, generating the three boundary divisors associated to the trivalent graphs $\gamma_s,\gamma_t,\gamma_u$, see figure~\ref{fig:sphere3}. 
We model the neighborhood of these degenerations by a thrice-punctured sphere $\Sigma_3$ with coordinate $z$ and the three marked points $z_s,z_t,z_u$ fixed to some values using the $\text{SL}(2,\C)$ symmetry. Near these degenerations the differential form $\varphi_+$ restricts to a 1-form on $\Sigma_3$ with poles along the marked points. Kinematical numerators are given by residues of this form
\begin{equation}
n_{\gamma_a} = \Res{z=z_a}\varphi_+\,,\qquad a\in\{s,t,u\}\,.
\end{equation}
and are related by a linear identity due to the global residue theorem on $\Sigma_3$,
\begin{equation}
n_{\gamma_s}+n_{\gamma_t}+n_{\gamma_u}=0\,.
\end{equation}
% The general argument can be illustrated on $\mathcal{M}_{0,4}$. Using the $\text{SL }(2,\C)$ symmetry we set three puncutures to fixed values $\{z_2,z_3,z_4\}$, the fourth puncture $z_1$ is allowed to vary and we identify $\mathcal{M}_{0,4}$ with a sphere with three punctures. In this case the numerator $\mathcal{I}$ is a one form with possible poles only when $z_1$ approaches one of the punctures. Near these points we can take its residues
% \begin{equation}
% \begin{array}{ccc}
%  \Res{z_1\rightarrow z_2}\mathcal{I}=\mathcal{I}_s \;\;\;&  \Res{z_1\rightarrow z_3}\mathcal{I}=\mathcal{I}_t \;\;\;&  \Res{z_1\rightarrow z_4}\mathcal{I}=\mathcal{I}_u
%  \end{array}
% \end{equation}
% and by the global residue theorem on the three-punctured sphere these are constrained to obey
% \begin{equation}
%  \mathcal{I}_s+\mathcal{I}_t+\mathcal{I}_u=0.
% \end{equation}
While this argument would work for any regular differential forms on $\mathcal{M}_{0,n}$, only forms which are elements of the twisted cohomology groups actually generate field theory amplitudes. Moreover, many simplifications occur by choosing good representatives for these cohomologies.

%%%%%%%%%%%%%%%%%%%%%%%%%%%%%%%%%%%%%
%%%%%%%%%%%%%%%%%%%%%%%%%%%%%%%%%%%%%

\section{Celestial scattering equations}
\label{sec:cse} 

The CHY formulas reviewed in the previous section have a natural celestial analogue. These are most straightforwardly -- if somewhat formally -- derived by directly Mellin transforming the momentum space expressions. We start this section by performing these Mellin transforms and writing down explicit formulas for celestial amplitudes. These take the form of Gelfand A-hypergeometric functions in all dimensions, now governed by the \emph{celestial scattering equations}. Some example computations are also provided.

%%%%%%%%%%%%%%%%%%%%%%%%%%%%%%%%%%%%%

\subsection{Amplitude formulas}
\label{subsec:formulas}

The $n$-point celestial amplitude can be written down as a Mellin transform of the momentum space formula \eqref{CHY} \cite{Adamo:2019ipt},
\begin{multline}\label{chymel}
\cA_n = \int_{\cM_{0,n}}\frac{\d^{n}z}{\text{vol SL}(2,\C)}\int_{\R^n_+}\prod_{j=1}^n\frac{\d\omega_j}{\omega_j}\,\omega_j^{\Delta_j}\;\pprod_i\bar\delta\biggl(\sum_{j\neq i}\frac{s_i\,s_j\,\omega_i\,\omega_j\,q_i\cdot q_j}{z_i-z_j}\biggr)\\
\times\mathcal{I}(\{g,s\,\omega\,q,\epsilon,z\})\;\tilde{\mathcal{I}}(\{g,s\,\omega\,q,\tilde\epsilon,z\})\,\delta^D\left(\sum_{i=1}^n s_i\,\omega_i\,q_i\right).
\end{multline}
The primary trick to simplify this is to perform as many $\omega_j$-integrals as possible against the momentum conserving delta function. We can already do this at the level of \eqref{chymel}, but we will find it much more illuminating to first manipulate it a bit.

For this we make use of the momentum generators in Mellin variables. The action of the momentum operator $P_{\al\dal}$ on a function $A(k)$ of null momentum $k=s\,\omega\,q$ is given by a trivial multiplication
\begin{equation}
P_{\mu}\cdot A(k) = k_{\mu}\,A(k)\,.
\end{equation}
Its action on the Mellin transform of $A(k)$ is then easily expressed as
\begin{equation}
P_{\mu}\cdot\int_{\R_+}\frac{\d\omega}{\omega}\,\omega^{\Delta}\,A(k) = s\,q_{\mu}\int_0^\infty\frac{\d\omega}{\omega}\,\omega^{\Delta+1}\,A(k) =  s\,q_{\mu}\,\e^{\p_{\Delta}}\int_0^\infty\frac{\d\omega}{\omega}\,\omega^{\Delta}\,A(k)\,.
\end{equation}
This dictates the definition of the celestial translation symmetry generators \cite{Stieberger:2018onx},
\begin{equation}\label{Kdef}
\cK_{\mu} := s\,q_{\mu}\,\e^{\p_\Delta}\,,
\end{equation}
which are operator-valued null vectors, $\cK_i^2=0$. Momentum conservation is then equivalent to invariance under the diagonal translation generator, that is,  $n$-point celestial amplitudes should be annihilated by
\begin{equation}
\sum_{i=1}^n\cK_i^{\mu} = \sum_{i=1}^n s_i\,q_i^{\mu}\,\e^{\p_{\Delta_i}}\,.
\end{equation}
Indeed, acting with this on \eqref{chymel} produces a factor of $\sum_i s_i\,\omega_i\,q_i^{\al\dal}$ inside the Mellin transforms which vanishes by momentum conservation. So, at least whenever the Mellin transforms converge (or are understood distributionally), celestial amplitudes are invariant under diagonal translations.

Hence we can make the formal replacements,
\be\label{replace}
\omega_i\mapsto\e^{\p_{\Delta_i}}\,,\qquad k_i\mapsto\cK_i\,,
\ee  
inside the scattering equations and the CHY integrands. By converting them into operators we can take these objects outside the Mellin integrals. Moreover, the various $\cK_i$'s clearly commute with each other and there is no operator ordering ambiguity. We thus find the following expression,
\be\label{cel_chy}
\cA_n = \int_{\cM_{0,n}}\frac{\d^{n}z}{\text{vol SL}(2,\C)}\;\pprod_i\bar\delta(\cE_i)\;\mathcal{I}(\{g,\cK,\epsilon,z\})\;\tilde{\mathcal{I}}(\{g,\cK,\tilde\epsilon,z\})\;\cS_n(\{\Delta,q,s\})\,,
\ee
having defined the \emph{celestial scattering equations},
\be\label{cEi}
\cE_i := \sum_{j\neq i}\frac{\cK_i\cdot\cK_j}{z_i-z_j}\,,
\ee
and the Mellin transformed $n$-point contact diagram,
\be\label{Sn}
\cS_n := \int_{\R^n_+}\prod_{j=1}^n\frac{\d\omega_j}{\omega_j}\,\omega_j^{\Delta_j}\,\delta^D\left(\sum_{i=1}^n s_i\,\omega_i\,q_i\right)\,.
\ee
Rigorously speaking, replacements like \eqref{replace} are supposed to be made inside analytic functions. In section \ref{sec:ambitwistor}, we will justify performing these replacements inside the delta functions $\bar\delta(E_i)$ by deriving it from the ambitwistor worldsheet CFT.

$\cS_n$ is the contribution of a $\phi^n$ contact term to the $n$-point celestial amplitude of a theory of massless scalars $\phi$. In appendix~\ref{app:Mellin} we explicitly evaluate these contact diagrams for arbitrary multiplicity and dimension, citing here only the final expression for the celestial CHY formulas,
\begin{multline}\label{Anmain}
\cA_n =\int_{\cM_{0,n}}\frac{\d^{n}z}{\text{vol SL}(2,\C)}\;\pprod_i\bar\delta(\cE_i)\;\mathcal{I}(\{g,\cK,\epsilon,z\})\;\tilde{\mathcal{I}}(\{g,\cK,\tilde\epsilon,z\})\\
\times F_n(\{\Delta,q,s\})\int_0^\infty\frac{\d\omega}{\omega}\,\omega^{\sum_i\Delta_i-D}\,.
\end{multline}
In particular, the replacement~\eqref{replace} applied to the Parke-Taylor factor~\eqref{PT} and the Pfaffian~\eqref{Pf} provides the celestial equivalent of kinematical numerators in~\eqref{cel_chy} for biadjoint scalar, Yang-Mills and gravity celestial amplitudes. Using the celestial scattering equations, these take the form of operator-valued numerators acting on the same universal function 
\begin{equation}
F_n(\{\Delta,q,s\})\int_0^\infty\frac{\d\omega}{\omega}\,\omega^{\sum_i\Delta_i-D}\,. 
\end{equation}
The function $F_n(\{\Delta,q,s\})$ takes different forms depending on whether $n>D$  or $n\leq D$.

\paragraph{$\boldsymbol{n>D:}$} In this case, there are more $\omega_j$ integrals than momentum conserving delta functions. This allows us to perform $D$ of the Mellin integrals -- say those over $\omega_1,\dots,\omega_D$ -- and the rest simplify to
\begin{multline}\label{Fn>D}
F_n(\{\Delta,q,s\}) = \frac{1}{u}\prod_{l,r}\Theta(u_{lr})\\
\times\int_{[0,1]^{n-D}}\prod_r\frac{\d\xi_r}{\xi_r}\,\xi_r^{\Delta_r}\,\prod_l\biggl(\sum_{r'}u_{lr'}\,\xi_{r'}\biggr)^{\Delta_l-1}\delta\biggl(1-\sum_{r''}\xi_{r''}\biggr)\,.
\end{multline}
Here, the various indices run over $l=1,2,\dots,D$ while $r,r',r''=D+1,\dots,n$. The coefficients $u$ and $u_{lr}$ in the integrand can be expressed in terms of $D\times D$ minors of the $D\times n$ matrix $(q_1^\mu, q_2^\mu,\dots,q_n^\mu)$ of celestial positions. Define the determinants,
\be
(i_1\,i_2\dots\,i_D) := \varepsilon_{\mu_1\mu_2\dots\mu_D}\,q_{i_1}^{\mu_1}\,q_{i_2}^{\mu_2}\cdots q_{i_D}^{\mu_D}\,,
\ee
where $\varepsilon_{\mu_1\mu_2\dots\mu_D}$ is the $D$-dimensional Levi Civita symbol. Then we can express them quite compactly as
\be
u = |(1\,2\dots D)|\,,\qquad u_{lr} = -s_l\,s_r\,\frac{(1\,2\dots l-1\;r\;l+1\dots D)}{(1\,2\dots D)}\,.
\ee
%\enote{I don't understand this definition}
The Heaviside step functions constrain the $u_{lr}$ to be positive. The integrals over the $\xi_r$'s produce Aomoto-Gelfand hypergeometric functions over the Grassmannian Gr$(n-D,n)$ (or equivalently on Gr$(D,n)$) \cite{Aomoto, Abe:2015ucn}. Such integrals were first identified in the context of celestial amplitudes in \cite{Schreiber:2017jsr}.

\paragraph{$\boldsymbol{n\leq D:}$} In this case, the result is distributional with $D-n+1$ leftover delta functions,
\be\label{FnlessD}
F_n(\{\Delta,q,s\}) = \delta^{D-n+1}\biggl(\sum_l s_l\,u_{ln}\,q_l + s_n\,q_n\biggr)\;\frac{1}{u}\prod_l\Theta(u_{ln})\,u_{ln}^{\Delta_l-1}\,.
\ee
Here, $l=1,2,\dots,n-1$. The spacetime index $\mu$ has also been partitioned into two sets: $\mu = (r,a)$, where $r=0,1,\dots,D-n$ while $a=D-n+1,\dots,D-1$. Then we have localized the Mellin integrals on the delta functions imposing $\sum_i s_i\,\omega_i\,q_i^a=0$ by solving for the $\omega_l$ in terms of $\omega_n$. The various coefficients in the above expression are
\be
u=|(1\,2\dots n-1)|\,,\qquad u_{ln} = -s_l\,s_n\,\frac{(1\,2\dots l-1\;n\;l+1\dots n-1)}{(1\,2\dots n-1)}\,,
\ee
where the determinants $(i_1\,i_2\dots i_{n-1})$ are now $(n-1)\times(n-1)$ minors of the $(n-1)\times n$ matrix $(q_1^{a}, q_2^{a},\dots,q_n^{a})$. Again, these coefficients satisfy positivity constraints $u_{ln}>0$ for all $l$. The remaining delta functions impose momentum conservation in the ``transverse'' $D-n+1$ dimensions. The result is still Lorentz invariant, the Mellin transforms themselves preserve the symmetry but in performing them explicitly, non-Lorentz covariant choices had to be made. 

We also remark that the leftover Mellin integral over $\omega$ in \eqref{Anmain} is generically divergent. An interpretation of such integrals was given in \cite{Donnay:2020guq} in terms of ``generalized delta functions'', allowing one to declare
\be\label{deltadef}
\int_0^\infty\frac{\d\omega}{\omega}\,\omega^{\sum_i\Delta_i-D} \equiv 2\pi\,\delta\!\left(\im\bigl(\textstyle\sum_i\Delta_i-D\bigr)\right)\,.
\ee
Formally, we see for instance from \eqref{FnlessD} that we indeed need the condition $\sum_i\Delta_i=D$ to hold so that it has conformal weight $\Delta_n$ in $q_n^\mu$. If the final amplitude is finite or marginally convergent, this awkwardness gets resolved by the application of the $\cK_i$'s on $\cS_n$ as this shifts the conformal weights appropriately. %\enote{we of course ends up with derivatives of delta functions. Might be wothwhile noting it, I think it also happens for loop amplitudes}. 
We discuss such shifts below.

\medskip

Let us now study some salient features of our formulas. First note that the various scattering equations $\cE_i$ trivially commute since the $\cK_i$ commute. Moreover, even though momentum conservation is absent, they are still $\text{SL}(2,\C)$ invariant when acting on translation invariant objects. This is a consequence of the fact that the residue of any $\cE_i$ at $z_i=\infty$ is given by
\be
\Res{z_i=\infty}\cE_i\,\d z_i = \cK_i\cdot\sum_{j=1}^n\cK_j\,,
\ee
were we used $\cK_i^2=0$. This residue is proportional to the diagonal translation generator which annihilates the contact diagram. 

Having evaluated the Mellin transform of $\cS_n$, we can use \eqref{Anmain} to solve the scattering equations and obtain the celestial S-matrix of a variety of theories. Since the operators $\cK_i$ formally behave as commuting numbers, the scattering equations are solved by the same algebraic expressions coming from the $(n-3)!$ solutions in momentum space. The worldsheet path integral implements a sum over all of these solutions and generates trivalent graphs $\gamma$ contributing to the $n$-point amplitude. The resulting celestial amplitude has the following general structure,
\begin{equation}\label{trigraph}
\cA_n = \sum_{\gamma\in\Gamma}\frac{\cN_\gamma\,\tilde\cN_\gamma}{\prod_{e\in\gamma}P_e}\cdot\cS_n\,.
\end{equation}
Each internal edge $e$ of a graph $\gamma$ now comes with an operator-valued propagator denominator,
\be
P_e = \left(\sum_{i\in e}\cK_i\right)^2 = 2\sum_{i,j\in e}s_i\,s_j\,q_i\cdot q_j\,\e^{\p_{\Delta_i}+\,\p_{\Delta_j}}\,.
\ee
The numerators $\cN_\gamma$ and $\tilde\cN_\gamma$ are also theory dependent operators given by residues of the CHY numerators $\mathcal{I}$ and $\tilde{\mathcal{I}}$ analogous to eq.~\eqref{residue}, see also section~\ref{sec:cel_ck}, which act on the celestial contact amplitude. 

For $n>D$, one can take these operators inside the Mellin integrals in $\cS_n$. For $i=r$, one straightforwardly replaces $\e^{\p_{\Delta_r}}\mapsto\xi_r$, while for $i=l$ one replaces $\e^{\p_{\Delta_l}}\mapsto \sum_r u_{lr}\,\xi_r$ as expected. Simultaneously, one also needs to appropriately shift $\sum_i\Delta_i$ occurring in the distributional factor \eqref{deltadef}, and this is most easily seen by working through the examples given in the next subsection. The resulting integrals yield Gelfand A-hypergeometric functions \cite{Gelfand, GKZ}. Similarly, for $n\leq D$, one can replace factors $\e^{\p_{\Delta_l}}\mapsto u_{ln}$ and $\e^{\p_{\Delta_n}}\mapsto 1$, while again shifting  $\sum_i\Delta_i$ appropriately.

%%%%%%%%%%%%%%%%%%%%%%%%%%%%%%%%%%%%%

\subsection{Examples}
\label{subsec:eg}

To illustrate our methods, we work out some examples of biadjoint scalar amplitudes in different dimensions. Generalizing to gluons and gravitons is computationally tedious but straightforward.

\paragraph{$\boldsymbol{4}$ points.} In this case, we only have one scattering equation. We choose to fix the three points $z_1,z_2,z_3$ so that the remaining scattering equation corresponds to $z_4$:
\begin{equation}
    \cE_4 \equiv  \frac{\cK_4\cdot\cK_1}{z_{41}} + \frac{\cK_4\cdot\cK_2}{z_{42}} + \frac{\cK_4\cdot\cK_3}{z_{43}} = 0\,.
\end{equation}
Fixing the global conformal symmetry also introduces a Faddeev-Popov determinant $(z_{12}z_{23}z_{31})^2$ into the worldsheet integrals, where $z_{ij}:=z_i-z_j$. Using $2\pi\im\,\bar\delta(\cE_i)=\dbar_i(1/\cE_i)$, we then need to evaluate
\be
\cA_4 = \oint\displaylimits_{\cE_4=0}\frac{\d z_4}{2\pi\im\,\cE_4}\;z_{12}^2\,z_{23}^2\,z_{31}^2\,\mathcal{I}\,\tilde{\mathcal{I}}\cdot\cS_4\,.
\ee

For convenience, we take the fixed points to be $z_1=0,z_2=1,z_3\to\infty$. As explained before, since the $\cK_i$'s commute, we can formally solve the scattering equations as in momentum space. We find
\begin{equation}
\frac{\cK_4\cdot\cK_1}{z_4} + \frac{\cK_4\cdot\cK_2}{z_4-1} = 0\implies z_4=z^*_4 = \frac{\cK_1\cdot\cK_4}{\cK_4\cdot(\cK_1+\cK_2)}\,.
\end{equation}
More precisely, this makes sense via its action $z^*_4\cdot\cS_4$ on the contact diagram which produces an ordinary number. To obtain particular amplitudes we can insert various different numerators. For example, we can compute the biadjoint scalar amplitude in the color ordering (1234|1234). This is done by substituting the (color-stripped) Parke-Taylor factors,
\be
\frac{1}{z_{12}\,z_{23}\,z_{34}\,z_{41}}\cdot\frac{1}{z_{12}\,z_{23}\,z_{34}\,z_{41}}\,,
\ee
for $\mathcal{I}\,\tilde{\mathcal{I}}$. Using $\sum_{i=1}^4\cK_i\cdot\cS_4 = 0$, one easily finds
\begin{equation}\label{A4bs}
     \begin{split}
         \cA_4(1234|1234) &= -\left(\frac{1}{\cK_1\cdot\cK_2} + \frac{1}{\cK_1\cdot\cK_4}\right)\cS_4\\
         &= -\left(s_1\,s_2\,\frac{\e^{-(\p_{\Delta_1}+\,\p_{\Delta_2})}}{q_1\cdot q_2} +s_1\,s_4\,\frac{\e^{-(\p_{\Delta_1}+\,\p_{\Delta_4})}}{q_1\cdot q_4}\right)\cS_4\,.
     \end{split}
\end{equation}
Clearly, all that the derivatives do is shift some of the weights by $-1$. This phenomenon is also clear from the perspective of the plane wave basis. There the propagator denominators contain factors of $\omega_i$'s which induce precisely these shifts.

Eq.~\eqref{A4bs} can be further simplified on each channel. When $D<4$, using \eqref{Fn>D} and \eqref{deltadef}, the $s$-channel contribution becomes
\begin{multline}
s_1\,s_2\,\frac{\e^{-(\p_{\Delta_1}+\,\p_{\Delta_2})}}{q_1\cdot q_2}\,\cS_4 = 2\pi\,\delta\!\left(\im\bigl(\textstyle\sum_{i=1}^4\Delta_i-D-2\bigr)\right)\frac{s_1\,s_2}{q_1\cdot q_2}\\
\times F_4(\Delta_1-1,q_1,s_1;\Delta_2-1,q_2,s_2;\Delta_3,q_3,s_3;\Delta_4,q_4,s_4)\,,
\end{multline}
with $F_4$ containing a Gr$(4-D,D)$ Aomoto-Gelfand hypergeometric integral. More interesting is the case $D\geq 4$. Here, one finds
\begin{multline}
s_1\,s_2\,\frac{\e^{-(\p_{\Delta_1}+\,\p_{\Delta_2})}}{q_1\cdot q_2}\,\cS_4 = 2\pi\,\delta\!\left(\im\bigl(\textstyle\sum_{i=1}^4\Delta_i-D-2\bigr)\right)\frac{s_1\,s_2}{q_1\cdot q_2}\\
\times\delta^{D-3}\biggl(\sum_{l=1}^3 s_l\,u_{l4}\,q_l + s_4\,q_4\biggr)\;\frac{u_{34}}{u}\prod_{l=1}^3\Theta(u_{l4})\,u_{l4}^{\Delta_l-2}\,.
\end{multline}
Other channels contribute similar terms.

\paragraph{Propagators at higher points.} 
At $n$ points, each Feynman diagram of the cubic biadjoint scalar theory comes with $n-3$ propagators and trivial numerators. Clearly, each propagator denominator induces a shift $\sum_i\Delta_i\mapsto\sum_i\Delta_i-2$ in the Mellin integral \eqref{deltadef}. Altogether, we find a factor,
\be\label{deltan}
\delta\!\left(\im\bigl(\textstyle\sum_{i=1}^n\Delta_i-D-2\,(n-3)\bigr)\right)\,,
\ee
in the final amplitude. With $D=d+2$, the on-shell phase space of massless particles in $\R^{1,D-1}$ is spanned by conformal basis states on the principal continuous series $\Delta_i\in\frac{d}{2}+\im\,\R$ \cite{Pasterski:2017kqt}.  So the distribution \eqref{deltan} is a true delta function precisely for $d=4$, i.e., in six dimensions. It is curious to note that $D=6$ is precisely the dimension in which the biadjoint scalar theory is a classical CFT.\footnote{Similarly, it is known from \cite{Pasterski:2017ylz} that Mellin transforms of gluon amplitudes are marginally convergent precisely for $D=4$: the dimension in which Yang-Mills is classically a CFT.}%\enote{I'd guess then that graviton amplitudes would give true delta functions in dimensions depending on $n$ following the weird conformal symmetry found by\cite{Loebbert:2018xce}}

In general, we will have propagator denominators acting on the celestial contact diagram. As a simple example, consider the action of a single propagator acting on $\cS_n$ for $n>D$,
\be
\frac{1}{(\sum_{i\in e}\cK_i)^2}\cdot\cS_n\,,
\ee
for an internal edge $e$ of a Feynman graph. When $n>D$, $F_n$ from \eqref{Fn>D} changes to
\begin{multline}
\frac{1}{u}\prod_{l,r}\Theta(u_{lr})\int_{[0,1]^{n-D}}\prod_r\frac{\d\xi_r}{\xi_r}\,\xi_r^{\Delta_r}\,\prod_l\biggl(\sum_{r'}u_{lr'}\,\xi_{r'}\biggr)^{\Delta_l-1}\delta\biggl(1-\sum_{r''}\xi_{r''}\biggr)\\
\times\left[\sum_{l,l'\in e}s_l\,s_{l'}\,q_l\cdot q_{l'}\sum_{r,r'} u_{lr}\,u_{l'r'}\,\xi_r\,\xi_{r'} + 2\sum_{l,r\in e}s_l\,s_r\,q_l\cdot q_r\sum_{r'}u_{lr'}\,\xi_r\,\xi_{r'} + \sum_{r,r'\in e}s_r\,s_{r'}\,q_r\cdot q_{r'}\,\xi_r\,\xi_{r'}\right]^{-1}\,.
\end{multline}
Due to the new quadratic polynomial in the integrand, the result now produces a Gelfand A-hypergeometric function \cite{Gelfand, GKZ}.

Every propagator is taken care of by insertion of such quadratic polynomials. We may also find the insertion of higher degree polynomials through the numerators in gluon and graviton amplitudes, which are also handled by the same theory of hypergeometric integrals. Similar integrals have also recently occurred in the physics of loop level Feynman integrals \cite{delaCruz:2019skx} as well as stringy canonical forms \cite{Arkani-Hamed:2019mrd}.

%\enote{this last bit needs some clean up}

%%%%%%%%%%%%%%%%%%%%%%%%%%%%%%%%%%%%%
%%%%%%%%%%%%%%%%%%%%%%%%%%%%%%%%%%%%%

\section{Ambitwistor strings}
\label{sec:ambitwistor}

%%%%%%%%%%%%%%%%%%%%%%%%%%%%%%%%%%%%%

\subsection{Models}

In this section, we derive our worldsheet formulas for celestial amplitudes using ambitwistor string theories~\cite{Mason:2013sva}. These are chiral string theories with target the space of null geodesics in Minkowski space. The worldsheet action of a large class of such models takes the general form,
\be\label{model}
S = \frac{1}{2\pi}\int_\Sigma P_\mu\dbar X^\mu - \mu\,T - e\,H + S_\text{matter}\,.
\ee
Here, $X^\mu$ and $P_\mu$ are fields of conformal weight $(0,0)$ and $(1,0)$ respectively on the string worldsheet $\Sigma$, and $S_\text{matter}$ is the action of a pair of auxiliary worldsheet CFTs. $T$ denotes the weight $(2,0)$ stress tensor, while $\mu$ and $e$ are weight $(-1,1)$ Beltrami differentials gauging the constraints $T=0$ and $H:=P^2/2=0$ respectively. The latter generates the gauge symmetry
\be\label{Hsym}
\delta X^\mu = \alpha P^\mu\,,\qquad\delta P_\mu = 0\,,\qquad\delta e = \dbar\alpha\,,
\ee
where $\alpha$ is a weight $(1,0)$ vector field on $\Sigma$. Thus, the constraint  $P^2=0$ and associated gauge redundancy $X\sim X + \alpha P$ reduce our target space to the space of null geodesics: ambitwistor space.

The choice of matter action $S_\text{matter}$ gives rise to amplitudes of a variety of theories:
\begin{align}
\text{Biadjoint scalar} &: \quad S_\text{matter} = S_\g + S_{\tilde\g}\,,\\
\text{Yang-Mills} &:\quad S_\text{matter} = S_\g + \frac{1}{2\pi}\int_\Sigma\frac{1}{2}\,\psi_\mu\dbar\psi^\mu-\chi\,\psi_\mu P^\mu\,,\\
\text{Gravity} &:\quad S_\text{matter} = \frac{1}{2\pi}\int_\Sigma\frac{1}{2}\,\psi_\mu\dbar\psi^\mu + \frac{1}{2}\,\tilde\psi_\mu\dbar\tilde\psi^\mu -\chi\,\psi_\mu P^\mu - \tilde\chi\,\tilde\psi_\mu P^\mu\,.
\end{align}
$S_\g$ and $S_{\tilde\g}$ denote actions of current algebra CFTs corresponding to Lie algebras $\g$ and $\tilde\g$. We will denote their weight $(1,0)$ currents as $j^\msf{a}$ and $\tilde\jmath^{\tilde{\msf{a}}}$ respectively. In the Yang-Mills and gravitational cases, the fields $\psi^\mu$, $\tilde\psi^\mu$ both denote weight $(\frac{1}{2},0)$ worldsheet fermions, while $\chi$ and $\tilde\chi$ are weight $(-\frac{1}{2},1)$ fermionic Lagrange multipliers gauging fermionic worldsheet symmetries akin to supersymmetry. Worldsheet correlators of the current algebra systems give rise to the Parke-Taylor type numerators \eqref{PT}, while those of the fermions produce the Pfaffian type numerators \eqref{Pf}.

Since both the $XP$ system and matter actions are free CFTs, it is easy to find their fundamental OPEs. In the former case, one finds
\be\label{XPope}
X^\mu(z)\,P_\nu(w) \sim \frac{\delta_\nu^\mu}{z-w}\,.
\ee
Similarly, the fermionic OPEs read,
\be\label{psiope}
\psi^\mu(z)\,\psi^\nu(w) \sim \frac{\eta^{\mu\nu}}{z-w}\,,\qquad\tilde\psi^\mu(z)\,\tilde\psi^\nu(w) \sim \frac{\eta^{\mu\nu}}{z-w}\,.
\ee
Lastly, the current algebra OPEs are the standard ones,
\be\label{jjope}
j^\msf{a}(z)\,j^\msf{b}(w)\sim \frac{f^\msf{abc}}{z-w}\,,\qquad \jt^{\tilde{\msf{a}}}(z)\,\jt^{\tilde{\msf{b}}}(w)\sim \frac{f^{\tilde{\msf{a}}\tilde{\msf{b}}\tilde{\msf{c}}}}{z-w}\,,
\ee
where $f^\msf{abc}$ and $f^{\tilde{\msf{a}}\tilde{\msf{b}}\tilde{\msf{c}}}$ are the structure constants of $\g$ and $\tilde\g$ respectively. The levels of the current algebras can be non-zero, producing multi-trace amplitudes, but since we're only going to be interested in the single trace contributions we omit terms proportional to the level.

As in standard string theory, on gauge fixing we will also add ghost fields for each of the gauge symmetries. For the bosonic symmetries generated by $T$ and $H$, one adds two $bc$ ghost systems consisting of fermionic ghosts $b,\tilde b$ with weights $(2,0)$ and $c,\tilde c$ with weights $(-1,0)$. For the supersymmetries generated by $\psi\cdot P$ and $\tilde\psi\cdot P$, one adds $\beta\gamma$ systems with bosonic ghosts $\beta,\tilde\beta$ of weights $(\frac{3}{2},0)$ and $\gamma,\tilde\gamma$ of weights $(-\frac{1}{2},0)$. These fields also have the well-known OPEs,
\be\label{bcope}
b(z)\,c(w) \sim \frac{1}{z-w}\,,\qquad \tilde b(z)\,\tilde c(w) \sim \frac{1}{z-w}\,,
\ee
and
\be\label{betaope}
\beta(z)\,\gamma(w) \sim \frac{1}{z-w}\,,\qquad \tilde\beta(z)\,\tilde\gamma(w) \sim \frac{1}{z-w}\,.
\ee
The expression for the BRST operator will vary depending on the matter content of the ambitwistor string. For the tree models we discuss, biadjoint scalar, Yang-Mills and gravity, their BRST operators square to zero with appropriate choices of dimensions and current algebra level. When setting up the model we implicitly assume to be working with choices that render the BRST charge nilpotent, but our final expressions after all the worldsheet calculations have been performed are actually valid in any spacetime dimension. They are after all, tree-level amplitudes of field theories, which are the same independent of the spacetime dimension. More details on the BRST charges can be found in~\cite{Mason:2013sva}.
% Ghosts will appear when writing the gauge fixed actions and vertex operators.\

% \medskip

Next, we construct vertex operators for these theories in the conformal primary basis of states. It is easiest to start with the biadjoint scalar states. Their fixed vertex operators are given by
\be\label{bsV}
V_i = c\,\tilde c\,j\cdot\msf{T}\,\jt\cdot\tilde{\msf{T}}\,\phi_i(X)\,,
\ee
where $i$ is a particle label and $\phi_i(X)\equiv\phi_{\Delta_i}^{s_i}(X;q_i)$ denotes the scalar conformal primary wavefunctions of \eqref{wavefunction_s}. Generators of the Lie algebras $\g$ and $\tilde \g$ are denoted by $\msf{T^a}$ and $\tilde{\msf{T}}^{\tilde{\msf{a}}}$ respectively. Integrated vertex operators are obtained form the usual descent procedure.
% , and we come to this point when we evaluate correlators of these operators.

Similarly, the fixed vertex operator of a conformal primary gluon external state in Yang-Mills is
\be
V_i^{-1} = c\,\tilde c\,\delta(\gamma)\,j\cdot\msf{T}\,\eps_i\cdot\psi\,\phi_i(X)\,,
\ee
where the superscript $-1$ stands for picture number. To construct the corresponding picture number 0 vertex operator, we descend by computing the OPE of $V_i^{-1}$ with the picture changing operator $\Upsilon = \delta(\beta)\,\psi\cdot P$. Using \eqref{XPope} and the scalar wavefunction \eqref{wavefunction_s}, it is easily seen that
\be\label{Pphiope}
P_\mu(z)\,\phi_i(X(w)) \sim \frac{\cK_{i\,\mu}\,\phi_i(X(w))}{z-w}\,.
\ee
Such OPEs are the means by which the celestial translation generators $\cK_{i\,\mu} = s_i\,q_{i\,\mu}\,\e^{\p_{\Delta_i}}$ of \eqref{Kdef} will enter our analysis. They act on the wavefunctions to their right. This produces the picture number 0 vertex operators,
\be
V_i^0 = c\,\tilde c\,j\cdot\msf{T}\left(\eps_i\cdot P + \eps_i\cdot\psi\,\cK_i\cdot\psi\right)\phi_i(X)\,.
\ee
A similar analysis holds for gravitons, yielding the conformal basis vertex operators,
\be
V_i^{-1,-1} = c\,\tilde c\,\delta(\gamma)\,\delta(\tilde\gamma)\,\eps_i\cdot\psi\,\eps_i\cdot\tilde\psi\,\phi_i(X)\,,
\ee
at picture number $-1$, and
\be
V_i^{0,0} = c\,\tilde c\left(\eps_i\cdot P + \eps_i\cdot\psi\,\cK_i\cdot\psi\right)\bigl(\tilde\eps_i\cdot P + \tilde\eps_i\cdot\tilde\psi\,\cK_i\cdot\tilde\psi\bigr)\,\phi_i(X)\,,
\ee
at picture number 0. We reiterate that, just like ordinary momenta, the $\cK_i$'s commute with each other and there is no ordering ambiguity here. BRST closure of these operators follows immediately from the conformal primary representatives being on-shell.

%%%%%%%%%%%%%%%%%%%%%%%%%%%%%%%%%%%%%

\subsection{Worldsheet correlators}

We start this section by computing the correlators of the biadjoint scalar vertex operators \eqref{bsV} in detail and deriving the celestial scattering equations. After this we give the generalization to Yang-Mills and gravity. 

To compute $n$-particle amplitudes we quantize these models on a $n$-punctured Riemann sphere with punctures located at $z_1,\dots,z_n$. We work in conformal gauge $\mu=0$ and also gauge fix $e$ to be an element of the $(n-3)$-dimensional Dolbeault cohomology group $H^{0,1}(\Sigma, T_\Sigma(z_1+\cdots+z_n))$. The gauge freedom \eqref{Hsym} is precisely the freedom in choosing a representative of its cohomology class. Explicitly, we pick the gauge fixing condition,
\be\label{efix}
e = \sum_{i=4}^n r_i\,e_i\,,
\ee
having chosen the punctures $z_4,\dots,z_n$ as our moduli without loss of generality. The $\{e_i\}$ denote a standard basis of this cohomology group. They act against quadratic differentials like $H$ by picking $2\pi\im$ times their residues at the $z_i$. So for instance we will find an insertion of 
\be\label{eHval}
\exp\left(-\frac{1}{2\pi}\int_\Sigma e\,H\right) = \exp\left(-\im\sum_{i=4}^nr_i\,\Res{z=z_i}H(z)\right)
\ee
inside the path integral for any correlator, coming from the action \eqref{model}. The $r_i$ provide $n-3$ moduli that are left to be integrated over after the gauge fixing.

The gauge fixed action of the biadjoint scalar ambitwistor string is given by
\begin{equation}\label{bsgf}
    S = \frac{1}{2\pi}\int_\Sigma P\cdot\dbar X + b\,\dbar c + \tilde b\,\dbar\tilde c + S_\g + S_{\tilde\g}\,.
\end{equation}
Using this action, the $n$-particle celestial amplitude is computed by the correlator,
\be\label{cAncor}
\cA_n = \int\displaylimits_{\Gamma\subset T^*\cM_{0,n}}\d^{n-3}z\;\d^{n-3}r\;\left\la\e^{-\frac{1}{2\pi}\int_\Sigma e\,H}\;\prod_{i=4}^n B_i\,\tilde B_i\;\prod_{j=1}^n V_j\right\ra\,,
\ee
where the integral is performed over an appropriate middle-dimensional contour $\Gamma$ in $T^*\cM_{0,n}$ \cite{Ohmori:2015sha}.\footnote{The exact definition of this contour will not be needed in what follows.} In this expression, we have inserted a product of $n-3$ picture changing operators,
\be
B_i = \frac{1}{2\pi}\int_\Sigma e_i\,b\,,\qquad \tilde B_i = \frac{1}{2\pi}\int_\Sigma e_i\,\tilde b\,,
\ee
needed to soak up fermionic zero modes and produce the measure on $\mathcal{M}_{0.n}$. Their OPEs with fixed vertex operators produce integrated vertex operators corresponding to a choice of $n-3$ $z_i$'s. The other three puncture locations $z_1,z_2,z_3$ have been fixed using the residual $\text{SL}(2,\C)$ symmetry.

One computes the products of $B_i$ and $\tilde B_i$ with $V_i$ using the OPEs \eqref{bcope} to find
\be
B_i\,\tilde B_i\cdot V_i = j\cdot\msf{T}\,\jt\cdot\tilde{\msf{T}}\,\phi_i(X(z_i))\,.
\ee
This strips off the $c$ and $\tilde c$ ghosts from $n-3$ of the vertex operators, and correlators of the remaining ghost zero modes produce a factor of $(z_{12}z_{23}z_{31})^2$ that can be accommodated by inserting a factor of $1/\vol\,\SL(2,\C)^2$ in the sense of Faddeev-Popov. Next, the current algebra correlators generate a pair of Parke-Taylor numerators PT$_n$ and $\widetilde{\mathrm{PT}}_n$ of the form \eqref{PT} ($\widetilde{\mathrm{PT}}_n$ contains traces over products of $\tilde{\msf{T}}^{\tilde{\msf{a}}_i}$). We throw out the multitrace terms since we're interested only in amplitudes where Yang-Mills states are exchanged. All these simplifications leave us with a correlator of the $XP$ CFT,
\be\label{cAncor1}
\cA_n = \int_{\Gamma}\frac{\d^nz\;\d^nr}{\vol\,\SL(2,\C)^2}\,\mathrm{PT}_n\;\widetilde{\mathrm{PT}}_n\;\left\la\e^{-\frac{1}{2\pi}\int_\Sigma e\,H}\;\prod_{j=1}^n\phi_i(X(z_i))\right\ra_{XP}\,.
\ee
We evaluate the last correlator by utilizing the OPE \eqref{Pphiope} computed before.

It is easily shown that $H(z) = \frac{1}{2}P^2(z)$ acts on a product of scalar wavefunctions $\phi_i(X)$ by the OPE,
\be
H(z)\,\prod_{i=1}^n\phi_i(X(z_i)) \sim \sum_j\sum_{k\neq j}\frac{\cK_j\cdot\cK_k}{(z-z_j)(z-z_k)}\;\prod_{i=1}^n\phi_i(X(z_i))\,.
\ee
As a result, on performing the $XP$ path integral, $H(z)$ is frozen to its ``classical'' value,
\be
H(z) = \sum_i\sum_{j\neq i}\frac{\cK_i\cdot\cK_j}{(z-z_i)(z-z_j)}\,.
\ee
Note that the double poles dropped out in this computation due to $\cK_i^2=0$. Finally, the exponential $\e^{-\frac{1}{2\pi}\int_\Sigma e\,H}$ can be brought outside the correlator. Using \eqref{eHval}, the $r$-integrals subsequently give rise to the $n-3$ scattering equations,
\be
\int\frac{\d^{n}r}{\vol\,\SL(2,\C)}\;\exp\left(-\im\sum_{i=4}^nr_i\,\Res{z=z_i}H(z)\right) = \pprod_i\bar\delta(\cE_i)\,,
\ee
where
\be
\Res{z=z_i}\sum_i\sum_{j\neq i}\frac{\cK_i\cdot\cK_j}{(z-z_i)(z-z_j)} = \sum_{j\neq i}\frac{\cK_i\cdot\cK_j}{z_i-z_j} \equiv \cE_i\,.
\ee
This justifies the replacements $\omega_i\mapsto\e^{\p_{\Delta_i}}$ done within the scattering equations in section~\ref{sec:cse}. The result is a top-form integrated over $\cM_{0,n}$.

The remaining correlator over the $XP$ system contains only the insertion $\prod_i\phi_i(X(z_i))$. Performing the path integral over $P_\mu$ imposes its equation of motion following from the gauge fixed action \eqref{bsgf}, $\dbar X^\mu = 0$. On $\Sigma = \CP^1$, this reduces the path integral over $X^\mu$ to an integral over its constant zero mode. Denoting this zero mode again by $X^\mu$, we find
\be
\left\la\prod_{i=1}^n\phi_i(X(z_i))\right\ra_{XP} = \int_{\R^{1,D-1}}\d^D X\;\prod_{i=1}^n\phi_i(X) \equiv \cS_n\,.
\ee
This is precisely the scalar contact celestial amplitude of \eqref{Sn}, as can be seen by expressing the scalar conformal primary wavefunctions in terms of momentum eigenstates via \eqref{wavefunction_s}. We thus arrive at the CHY formula for celestial amplitudes of the biadjoint scalar theory,
\be\label{Anbs}
\cA_n =\int_{\cM_{0,n}}\frac{\d^{n}z}{\text{vol SL}(2,\C)}\;\pprod_i\bar\delta(\cE_i)\;\mathrm{PT}_n\;\widetilde{\mathrm{PT}}_n\;\cS_n(\{\Delta,q,s\})\,.
\ee
Translation invariance of $\cS_n$ and commutativity of the $\cK_i$'s guarantees permutation invariance of the scattering equations, so all choices of $n-3$ scattering equations are equivalent.

\medskip

An analogous story holds for Yang-Mills and gravitational amplitudes. For instance, in the former case we need to compute a correlator of the form,
\be\label{cAncorym}
\cA_n =\int\displaylimits_{\Gamma\subset T^*\cM_{0,n}}\d^{n-3}z\;\d^{n-3}r\;\left\la\e^{-\frac{1}{2\pi}\int_\Sigma e\,H}\;\prod_{i=4}^n B_i\,\tilde B_i\;V_1^{-1}\,V_2^{-1}\,\prod_{j=3}^n V_j^0\right\ra\,,
\ee
having chosen to insert picture number $-1$ vertex operators for particles 1 and 2 without loss of generality. The path integrals over the fermionic ghosts is done as usual and we pull out the insertion of $\e^{-\frac{1}{2\pi}\int_\Sigma e\,H}$ from the correlator as before leading to the celestial scattering equations. The path integral over the currents $j^\msf{a}$ generates a single Parke-Taylor factor plus multitrace terms we once again ignore. This leaves the correlators of the $XP$ and $\psi$ systems along with the $\beta\gamma$ ghosts to be performed,
\be
\la\delta(\gamma(z_1))\,\delta(\gamma(z_2))\ra_{\beta\gamma}\left\la\prod_{i=3}^n\left(\eps_i\cdot P + \eps_i\cdot\psi\,\cK_i\cdot\psi\right)\!(z_i)\;\prod_{j=1}^n\phi_i(X(z_j))\right\ra_{XP\psi}\,.
\ee
Following~\cite{Mason:2013sva}, these yield a Pfaffian-type numerator \eqref{Pf}, now containing the operators $\cK_i$ in place of ordinary momenta $k_i$. The final result is the formula,
\be\label{Anym}
\cA_n =\int_{\cM_{0,n}}\frac{\d^{n}z}{\text{vol SL}(2,\C)}\;\pprod_i\bar\delta(\cE_i)\;\mathrm{PT}_n\;\mathrm{Pf}'\Psi_n(\{\eps,\cK\})\;\cS_n(\{\Delta,q,s\})\,,
\ee
encoding gluon celestial amplitudes in arbitrary dimensions.

For gravity, the only new ingredient is the replacement of the current algebra system with a second fermionic system. The correlator of interest is given by
\be\label{cAncorgr}
\cA_n =\int\displaylimits_{\Gamma\subset T^*\cM_{0,n}}\d^{n-3}z\;\d^{n-3}r\;\left\la\e^{-\frac{1}{2\pi}\int_\Sigma e\,H}\;\prod_{i=4}^n B_i\,\tilde B_i\;V_1^{-1,-1}\,V_2^{-1,-1}\,\prod_{j=3}^n V_j^{0,0}\right\ra\,.
\ee
Using the same calculations as above it reduces to the graviton celestial amplitude formula
\be\label{Angr}
\cA_n =\int_{\cM_{0,n}}\frac{\d^{n}z}{\text{vol SL}(2,\C)}\;\pprod_i\bar\delta(\cE_i)\;\mathrm{Pf}'\Psi_n(\{\eps,\cK\})\;\mathrm{Pf}'\Psi_n(\{\tilde\eps,\cK\})\;\cS_n(\{\Delta,q,s\})\,.
\ee
There are two unifying features of all these formulas. The first is the presence of the same contact amplitude $\cS_n$ and celestial scattering equations governing the three expressions. The second is the manifest double copy structure at the level of the CHY integrands given by a simple replacement rule. In the next section, we return to this point with the machinery of twisted cohomology as applied to these operator-valued integrands.

%%%%%%%%%%%%%%%%%%%%%%%%%%%%%%%%%%%%%
%%%%%%%%%%%%%%%%%%%%%%%%%%%%%%%%%%%%%

\section{Celestial color-kinematics duality}
\label{sec:cel_ck}

We have seen in the previous section how the celestial versions of the CHY formulas still manifest a double copy structure for the \textit{operator-valued} numerators. This can either be stated as a substitution rule for ambitwistor integrands or in terms of their residues and trivalent graphs. In the latter method, gauge invariance of the double copied amplitude is not manifest even though we know it is gauge invariant by construction. As reviewed in section~\ref{sec:review} gauge invariance of the double copied amplitude is guaranteed if the kinematical numerators obey color-kinematics duality. We show below that an analogous requirement holds for the operator-valued numerators of celestial amplitudes obtained from the ambitwistor string. The proof will closely follow~\cite{Mizera:2019blq} by recasting the operator-valued numerators as elements of a generalized twisted cohomology.

% in we first set up a double copy analysis for celestial amplitudes that is completely extricated from the plane wave basis. To do this, we need to use the proof of color-kinematics duality that follows from Mizera's intersection theoretic analysis .

%\subsection{Twisted cohomology}
%\label{sec:op_t_c}
\medskip

Let $\cC_n$ denote the configuration space for the celestial data $\{\Delta_i,q_i,s_i\}$ of $n$-particle celestial amplitudes. The operators $\cK_i$ defined in~\eqref{Kdef} map smooth functions on $\cC_n$ to themselves by shifting their conformal dimensions. These operators are not derivations as they don't satisfy Leibniz rule, but they are linear and commute with each other. Denote the $C^\infty(\cC_n)$-algebra generated by the $\cK_i$'s by $\mathfrak{C}$. It forms a subalgebra of $\mathrm{Hom}(C^\infty(\cC_n),C^\infty(\cC_n))$. Even though translation invariance is not manifest it is still a symmetry of celestial amplitudes, so our operators $\cK_i$ will only ever act on functions $\mS$ satisfying
\begin{equation}
\sum_{i=1}^n\cK_i\mS = 0\,.
\end{equation}
Without loss of generality, we quotient $\mfk{C}$ by the ideal generated by $\sum_i\cK_i$ to define
\begin{equation}
\mfk{K} := \mfk{C}/\la\textstyle{\sum}_i\cK_i\ra\,
\end{equation}
as the reduced kinematical space where celestial amplitudes live. In what follows, we will treat the operator-algebras $\mfk{C}$ and $\mfk{K}$ as infinite-dimensional vector spaces spanned by monomials in $\cK_i$'s. 

%With some abuse of notation, we also denote by $\mfk{K}$ the infinite-dimensional trivial vector bundle $\cM_{0,n}\times\mfk{K}$.\footnote{We are treating $\mfk{C}$ and $\mfk{K}$ as infinite-dimensional vector spaces spanned by monomials in $\cK_i$'s.}  In analogy with the usual twisted cohomology, we define $\mfk{K}$-valued twisted de Rham operators,
%\begin{equation}
%\nabla_\pm = \d \pm \omega\,,
%\end{equation}
%with $\d$ the exterior derivative on $\cM_{0,n}$ and the connection one-form built out of the celestial scattering equations,
%\begin{equation}
%\omega = \psum_i \mathcal{E}_i\,\d z_i\,,\qquad \mathcal{E}_i = \sum_{j\neq i}\frac{\cK_i\cdot\cK_j}{z_{ij}}\,.
%\end{equation}
%The operator-valued connections $\nabla_+$ and $\nabla_-$ act via straightforward multiplication on twisted forms valued in $\mfk{K}$, denoted by $\Omega^{\bullet}(\cM_{0,n},\mfk{K})$.\footnote{The images of $\sum_i\cK_i$ under $\nabla_\pm$ are again in $\la\sum_i\cK_i\ra$ since the $\cK_i$ operators multiply as usual, commute with each other and with $\d$.} These connections $\nabla_\pm$ again square to zero, so we use them to define twisted de Rham cohomologies with coefficients in $\mfk{K}$ denoted by $H_{\nabla_\pm}^\bullet(\cM_{0,n},\mfk{K})$.

With some abuse of notation, we also denote by $\mfk{K}$ the infinite-dimensional trivial vector bundle $\cM_{0,n}\times\mfk{K}$. In analogy with the usual twisted cohomology, we define the twisted de Rham operators
\begin{equation}
\nabla_\pm = \d \pm \omega\,,
\end{equation}
with $\d$ the exterior derivative on $\cM_{0,n}$ and the connection one-form built out of the celestial scattering equations,
\begin{equation}
\omega = \psum_i \mathcal{E}_i\,\d z_i\,,\qquad \mathcal{E}_i = \sum_{j\neq i}\frac{\cK_i\cdot\cK_j}{z_{ij}}\,.
\end{equation}
The operator-valued connections $\nabla_+$ and $\nabla_-$ act via straightforward multiplication on forms valued in $\mfk{K}$, denoted by $\Omega^{\bullet}(\cM_{0,n},\mfk{K})$.\footnote{The images of $\sum_i\cK_i$ under $\nabla_\pm$ are again in $\la\sum_i\cK_i\ra$ since the $\cK_i$ operators multiply as usual, commute with each other and with $\d$.} We also formally define the line bundles $\mfk{L}$ and $\mfk{L}^\vee$ whose sections are $\mfk{K}$-valued functions on $\cM_{0,n}$.\footnote{More precisely, $\mfk{L}$ and $\mfk{L}^\vee$ are local systems given by an abelian operator-valued representation of $\pi(\mathcal{M}_{0,n})$. We leave the study of analytic aspects of these definitions to future work.} The connections $\nabla_\pm$ square to zero so we use them to define twisted de Rham cohomologies with coefficients in $\mfk{L}$ and $\mfk{L}^\vee$, denoted by $H^\bullet(\cM_{0,n},\mfk{L})$ and $H^\bullet(\cM_{0,n},\mfk{L}^\vee)$.

The operators $\cK_i$, viewed as elements of $\mfk{K}$, obey the same algebraic identities as the usual momenta for plane waves $k_i$. That is, they are null, commute with each other and obey a ``momentum conservation'' identity. Due to this, the analysis in~\cite{Mizera:2019gea,Mizera:2019blq} carries over to the celestial case with the appropriate substitutions. %For generic external celestial data the only non-vanishing cohomology groups are the middle-dimensional ones, $H^{n-3}(\mathcal{M}_{0,n},\cL\otimes\mfk{K})$.
 Let $\varphi_\pm$ be two $\mfk{K}$-valued meromorphic top forms on $\cM_{0,n}$ with singularities only along its boundary divisor. Take these to also obey $\nabla_\pm\varphi_\pm=0$ so that they are representatives of cohomology classes in $H^{n-3}(\cM_{0,n},\mfk{L})$ and $H^{n-3}(\cM_{0,n},\mfk{L}^\vee)$. Explicit examples are the ambitwistor numerators \eqref{PT} and \eqref{Pf} with the replacement $k_i\mapsto\cK_i$, multiplied by the top holomorphic form $\d \mu$ on $\mathcal{M}_{0,n}$.

We define intersection numbers $\la\varphi_+|\varphi_-\ra$ in analogy with~\cite{Mizera:2019gea} but in our case this pairing is not a number but an operator. There are two common ways to compute these intersection numbers: one is given by the CHY formulas, as a moduli space integral localized to the solutions of the scattering equations; the other by evaluating the paired forms $\varphi_{\pm}$ near the highest codimension boundaries of $\mathcal{M}_{0,n}$\footnote{Explicitly, this requires a finding a compactly supported cohomological representative of one of the forms $\varphi_\pm$. This coincides with the formula obtained by deforming the contours given by the scattering equations.} labelled by trivalent graphs. The first method recovers the formula \eqref{Anmain} for numerators $\varphi_+=\mathcal{I}\,\d \mu$ and $\varphi_-=\tilde{\mathcal{I}}\,\d \mu$. The latter gives the amplitude as a sum over trivalent graphs $\gamma$,
\begin{equation}\label{dc}
\la\varphi_+|\varphi_-\ra = \sum_{\gamma\in\Gamma}\frac{\mathrm{Res}_{v_\gamma}(\varphi_+)\,\mathrm{Res}_{v_\gamma}(\varphi_-)}{\prod_{e\in\gamma}P_e}\,,
\end{equation}
with edges denoted by $e$. Residues are taken along the boundary divisor $v_\gamma$ associated to the trivalent graph $\gamma$. Each edge has an associated denominator $P_e = (\sum_{i\in e}\cK_i)^2$ analogous to propagators acting as inverse operators. This is nothing other than eq.~\eqref{trigraph} which was obtained by deforming the contours prescribed by the scattering equations.

With the framework introduced above it is straightforward to adapt the arguments of~\cite{Mizera:2019blq} to show how color-kinematics is implemented for celestial amplitudes. The set up is analogous to section~\ref{subsec:t_cohomology}. Take a triple of trivalent graphs $\gamma_s,\gamma_t,\gamma_u$ differing only on a subgraph connecting four edges as shown in figure~\ref{fig:BCJ_graphs}. The boundary divisors for these graphs arise from a four-punctured sphere degenerating as one of its punctures $z$ approaches one of the other punctures $z_s,z_t,z_u$. The neighborhood of this degeneration is modelled by a sphere with three punctures, $\Sigma_3$, with coordinate $z$ and fixed marked points $z_s,z_t,z_u$, see figure~\ref{fig:sphere3}. Take $\varphi_M$ as a $\mfk{K}$-valued 1-form on $\Sigma_3$ with poles along the marked points. The operator-valued numerators,
\begin{equation}
\cN_{\gamma_a} = \Res{z=z_a}\varphi_M\,,\qquad a\in\{s,t,u\}\,,
\end{equation}
are then related by a linear identity due to the global residue theorem on $\Sigma_3$,
\begin{equation}
\cN_{\gamma_s}+\cN_{\gamma_t}+\cN_{\gamma_u}=0\,,
\end{equation}
valid in $\mfk{K}$, i.e. the sum of operators on the left vanishes precisely when it acts on translationally invariant functions. In $\mfk{C}$, that is, before taking the quotient, the statement of color-kinematics duality is
\begin{equation}\label{kJid}
\cN_{\gamma_s}+\cN_{\gamma_t}+\cN_{\gamma_u}= \mathcal{O}\cdot\sum_{i=1}^n\cK_i,
\end{equation}
for some operator $\mathcal{O}^\mu\in\mfk{C}$. 

% For contact amplitudes, $\cN=0$.\footnote{The only other symmetries in general are the Lorentz transformations, but their celestial generators don't come with factors of $\e^{\p_{\Delta_i}}$'s and so don't belong to $\mfk{C}$.} 

\medskip

As an example, we illustrate \eqref{kJid} for the four-point Yang-Mills amplitude. The operator-valued numerators are easily read off from their momentum space counterparts. We have
\begin{multline}
\cN_s = \left[\eps_1\cdot\eps_2\,(\cK_1-\cK_2)^\mu + 2\,\eps_1\cdot\cK_2\,\eps_2^\mu-2\,\eps_2\cdot\cK_1\,\eps_1^\mu\right]\eta_{\mu\nu}\\
\times\left[\eps_3\cdot\eps_4\,(\cK_4-\cK_3)^\nu - 2\,\eps_3\cdot \cK_4\,\eps_4^\nu+2\,\eps_4\cdot \cK_3\,\eps_3^\nu\right]\\
-\left(\eps_1\cdot\eps_3\,\eps_2\cdot\eps_4-\eps_1\cdot\eps_4\,\eps_2\cdot\eps_3\right)\left(\cK_1\cdot \cK_2+\cK_3\cdot \cK_4\right)\,
\end{multline}
for the $s$-channel. Other channels are obtained from permutations of this: $\cN_t$ by exchanging $2\leftrightarrow3$ with an overall minus sign, and $\cN_u$ by the permutation $(2,3,4)\rightarrow(4,2,3)$. Making use of $q_i^2=\eps_i\cdot q_i=0$, these three numerators are easily shown to obey
\begin{equation}
\cN_s+\cN_t+\cN_u = \left(\mathcal{R}_{1234}^\mu + \mathcal{R}_{1342}^\mu + \mathcal{R}_{1423}^\mu\right)\sum_{i=1}^4\cK_{i\,\mu}\in\la\textstyle\sum_i\cK_i\ra\,,
\end{equation}
where
\begin{equation}
\mathcal{R}_{ijkl}^\mu = 2\,\bigl(\eps_i^\nu\,\eps_j^\mu-\eps_j^\nu\,\eps_i^\mu\bigr)\,(\cK_k-\cK_l)_\nu + 2\left(\eps_k^\nu\,\eps_l^\mu-\eps_l^\nu\,\eps_k^\mu\right)(\cK_i-\cK_j)_\nu\in\mfk{C}\,.
\end{equation}
As claimed, the kinematic Jacobi identity is violated only up to something proportional to the diagonal translation generators.

With this we have shown that numerators given by $\mathrm{Res}_{v_\gamma}(\varphi_\pm)$ satisfy the celestial color-kinematics duality and \eqref{dc} makes manifest the the celestial double copy structure for any number of external particles extending the results of~\cite{Casali:2020vuy}. 

%%%%%%%%%%%%%%%%%%%%%%%%%%%%%%%%%%%%%
%%%%%%%%%%%%%%%%%%%%%%%%%%%%%%%%%%%%%

\section{Discussion}
\label{sec:end}

Ambitwistor strings are important tools for the study of flat space holography as they provide a framework to study celestial amplitudes to any multiplicity and in several dimensions. This is exemplified by our all-multiplicity expressions \eqref{Anmain}, \eqref{Fn>D} and \eqref{FnlessD} which allowed us to generalize the computations of \cite{Schreiber:2017jsr} to general dimensions and polarizations. In fact, we showed that for multiplicities with $n>D$, amplitudes in $D$ dimensions can be uniformly expressed in terms of Gelfand A-hypergeometric functions. For $n\leq D$, we found distributional expressions with $(D-n+1)$-dimensional delta functions coming from residual momentum conservation. We also observed that the Mellin transforms of biadjoint scalar amplitudes are marginally convergent in $D=6$, just as the Yang-Mills ones in $D=4$ \cite{Pasterski:2017ylz}. These happen to be the dimensions in which the classical theories are conformal.

There are also worldsheet models specialized to $D=4$ with target space $\scri$ that have been shown to compute Yang-Mills and gravity amplitudes in the plane wave basis~\cite{Adamo:2014yya,Adamo:2015fwa}. These models are adapted to the spinor-helicity formalism and might be better adapted to the study of a conjectural $2$d CFT on $\scri$. It would be interesting to understand how the computational methods we used above can be adapted to this case, and to compute the celestial amplitudes in these $D=4$ models.

Beyond explicit expressions for the $n$-point celestial amplitudes, our worldsheet formulas have provided a new outlook on their double copy. Celestial color-kinematics duality and double copy depend on two properties of the amplitudes: kinematical numerators can be represented as operators acting on external kinematics, that is, acting only on quantities at the boundary of spacetime; and that total derivatives can also be characterized as operators acting only on the external kinematics. The latter is simply the statement that amplitudes are translation invariant in the celestial case even if not manifestly so. 

These properties might be expected to hold only in the simpler case of flat spacetime. But we know of at least one case where they also hold for amplitudes in a curved background, namely, AdS. The works \cite{Roehrig:2020kck, Eberhardt:2020ewh} used the ambitwistor string to write CHY-like formulas for amplitudes in AdS spacetimes. These have similar structure to the celestial ones, with kinematical numerators and the scattering equations acting as operators on the scalar contact vertex. These amplitudes are not translation invariant. Instead, the decoupling of total derivatives on the scalar contact diagram can be identified with a quotient by the ideal of diagonal conformal transformations in analogy with the celestial case. We then expect that a similar notion of color-kinematics duality holds for AdS with kinematical numerators obeying a relation like~\eqref{kJid} up to some symmetry of the space of external data.

Double copy in AdS can also be expected to hold in a similar fashion to the celestial double copy. To show this we must first find the appropriate CHY numerators taking into account that some terms which naively look like non-zero contributions might decouple on top of the scattering equations. To characterize such terms a generalization of twisted cohomology analogous to the one we defined for celestial amplitudes would be very useful. We expect that several acceptable numerators could be found using insights from the ambitwistor string together with an interpretation of these numerators as a generalized twisted cohomology.

An interesting question is whether the framework introduced above holds for loop amplitudes. There are a couple of ambitwistor formulas for loop amplitudes~\cite{Adamo:2013tsa,Adamo:2015hoa}, the most successful being the ones based on nodal surfaces~\cite{Geyer:2015bja,Geyer:2015jch,Geyer:2016wjx,Geyer:2018xwu}. The latter can be generalized to the celestial case making use of our replacement rule \eqref{replace} to extract numerators in front of the Mellin transform of a scalar quantity. The outstanding issue is that the loop-level scattering equations and the numerators can depend on the loop momentum which we'd rather not have in a purely celestial description. If one goes back to position space Feynman diagrams this problem is absent since we can leave all the internal propagators in their position space representation. Numerators that don't depend on loop momentum can still be pulled out in front as operators acting on a scalar loop diagram, but now there is a proliferation of different topologies coming from the loop diagrams. It would be interesting to see if there is a way to encode the effect of loop momentum in the numerators in terms of operators acting on the external variables. Perhaps, some insight could be gained from explicit Mellin transforms giving loop-level celestial amplitudes~\cite{Banerjee:2017jeg,Albayrak:2020saa,Gonzalez:2020tpi}.

%\enote{What else?}

\section*{Acknowledgements}

We thank Tim Adamo and Andrea Puhm for comments on the draft. EC's research is supported in part by U.S. Department of Energy grant DE-SC0009999 and by funds provided by the University of California. AS is supported by a Mathematical Institute Studentship, Oxford.

\begin{appendix}

\section{Mellin transform of the contact diagram}
\label{app:Mellin}

In this appendix we evaluate the celestial $n$-point contact diagram
\be
\cS_n = \int_{\R^n_+}\prod_{i=1}^n\frac{\d\omega_i}{\omega_i}\,\omega_i^{\Delta_i}\;\delta^D\biggl(\sum_{j=1}^n s_j\,\omega_j\,q_j\biggr)\,.
\ee
There are two cases to consider: $n>D$ and $n\leq D$. In the former case, we will find the structure of Aomoto-Gelfand hypergeometric integrals which also appear in four dimensions~\cite{Schreiber:2017jsr}. In the latter case, we will be able to perform all the Mellin integrals and the result will be distributional due to leftover delta functions.

\paragraph{a) $\boldsymbol{n>D:}$}We begin by solving the delta functions for $\omega_l$, $l=1,2,\dots,D$, in terms of $\omega_r$, $r=D+1,\dots,n$, and the other variables. To do this, we use Cramer's rule. Define the determinants
\be
(i_1\,i_2\dots i_D) := \varepsilon_{\mu_1\mu_2\dots\mu_D}\,q_{i_1}^{\mu_1}\,q_{i_2}^{\mu_2}\cdots q_{i_D}^{\mu_D}\,,
\ee
where $\varepsilon_{\mu_1\mu_2\dots\mu_D}$ stands for the $D$-dimensional Levi Civita symbol. Then first write
\be
\sum_l s_l\,\omega_l\,q_l^\mu = -\sum_r s_r\,\omega_r\,q_r^\mu\,.
\ee
Contracting both sides with $\veps_{\mu\mu_1\mu_2\dots\hat\mu_l\dots\mu_D}\,q_1^{\mu_1}\,q_2^{\mu_2}\cdots\widehat{ q_l^{\mu_l}}\cdots q_D^{\mu_D}$ (where a hat denotes omission), we can solve for $\omega_l$ to find
\be
\omega_l = \sum_r u_{lr}\,\omega_r\,,\qquad u_{lr} = -s_l\,s_r\,\frac{(1\,2\dots l-1\;r\;l+1\dots D)}{(1\,2\dots D)}\,.
\ee
Note that demanding $\omega_l>0$ for all $\omega_r>0$ restricts all the coefficients $u_{lr}$ to be positive. Accounting for this using heaviside step functions, along with a Jacobian $u=|(1\,2\dots D)|$ coming from solving the delta functions, our integral turns into
\be
\begin{split}
\cS_n &= \int_{\R^n_+}\prod_{i=1}^n\frac{\d\omega_i}{\omega_i}\,\omega_i^{\Delta_i}\;\frac{1}{u}\prod_l\left\{\delta\biggl(\omega_l - \sum_{r'} u_{lr'}\,\omega_{r'}\biggr)\prod_r\Theta(u_{lr})\right\}\\
&= \frac{1}{u}\prod_{l,r}\Theta(u_{lr})\int_{\R^{n-D}_+}\prod_r\frac{\d\omega_r}{\omega_r}\,\omega_r^{\Delta_r}\,\prod_l\left(\sum_{r'}u_{lr'}\,\omega_{r'}\right)^{\Delta_l-1}\,.
\end{split}
\ee
The leftover integral has the standard form of an Euler-type integral.

To perform the $\omega_r$ integrals, define new integration variables $\omega := \sum_r\omega_r$ along with $\xi_r := \omega_r/\omega$. Doing this, we find
\begin{multline}
\cS_n = \frac{1}{u}\prod_{l,r}\Theta(u_{lr})\int_0^\infty\frac{\d\omega}{\omega}\,\omega^{\sum_i\Delta_i-D}\\
\times\int_{[0,1]^{n-D}}\prod_r\frac{\d\xi_r}{\xi_r}\,\xi_r^{\Delta_r}\,\prod_l\left(\sum_{r'}u_{lr'}\,\xi_{r'}\right)^{\Delta_l-1}\delta\biggl(1-\sum_{r''}\xi_{r''}\biggr)\,.
\end{multline}
The integral over $\omega$ is divergent but is usually interpreted as a distribution \cite{Donnay:2020guq}. Note that the degree of divergence will change on application of the $\cK_i$ operators in the scattering equations and kinematical numerators. The integral over the $\xi_r$'s has the structure of an Aomoto-Gelfand hypergeometric function.

\paragraph{b) $\boldsymbol{n\leq D:}$} In this case, the delta functions can be used to perform all the Mellin transforms. To do this, we solve for $\omega_l$, $l=1,\dots,n-1$, in terms of $\omega_n$. Divide the spacetime index $\mu$ into two sets: $\mu = (r,a)$, where $r=0,1,\dots,D-n$ while $a=D-n+1,\dots,D-1$. We solve the last $n-1$ momentum conservation constraints,
\be
\sum_ls_l\,\omega_l\,q_l^{a} = -s_n\,\omega_n\,q_n^{a}\,,
\ee
for the $\omega_l$. This yields
\be
\omega_l = u_{ln}\,\omega_n\,,\qquad u_{ln} = -s_l\,s_n\,\frac{(1\,2\dots l-1\;n\;l+1\dots n-1)}{(1\,2\dots n-1)}\,,
\ee
where the determinants $(i_1\,i_2\dots i_{n-1})$ are now defined as
\be
(i_1\,i_2\dots i_{n-1}) := \varepsilon_{a_1a_2\dots a_{n-1}}\,q_{i_1}^{a_1}\,q_{i_2}^{a_2}\cdots q_{i_{n-1}}^{a_{n-1}}\,,
\ee
with $ \varepsilon_{a_1a_2\dots a_{n-1}}$ the $(n-1)$-dimensional Levi Civita symbol. Again, these solutions are supplemented by the constraints $u_{ln}>0$ for all $l$.

Renaming $\omega_n\equiv\omega$, we find the expression
\be
\cS_n = \delta^{D-n+1}\biggl(\sum_l s_l\,u_{ln}\,q_l + s_n\,q_n\biggr)\;\frac{1}{u}\prod_l\Theta(u_{ln})\,u_{ln}^{\Delta_l-1}\int_0^\infty\frac{\d\omega}{\omega}\,\omega^{\sum_i\Delta_i-D}\,,
\ee
where the Jacobian is now $u=|(1\,2\dots n-1)|$. The remaining delta functions constrain the $q_i^r$ for $r=0,1,\dots,D-n$. 

\end{appendix}

\bibliographystyle{JHEP}
\bibliography{cdc}

\providecommand{\href}[2]{#2}\begingroup\raggedright\begin{thebibliography}{10}

\bibitem{Cheung:2016iub}
C.~Cheung, A.~de~la Fuente, and R.~Sundrum, {\it {4D scattering amplitudes and
  asymptotic symmetries from 2D CFT}},  {\em JHEP} {\bf 01} (2017) 112,
  [\href{http://arxiv.org/abs/1609.00732}{{\tt arXiv:1609.00732}}].

\bibitem{Pasterski:2016qvg}
S.~Pasterski, S.-H. Shao, and A.~Strominger, {\it {Flat Space Amplitudes and
  Conformal Symmetry of the Celestial Sphere}},  {\em Phys. Rev. D} {\bf 96}
  (2017), no.~6 065026, [\href{http://arxiv.org/abs/1701.00049}{{\tt
  arXiv:1701.00049}}].

\bibitem{Pasterski:2017kqt}
S.~Pasterski and S.-H. Shao, {\it {Conformal basis for flat space amplitudes}},
   {\em Phys. Rev. D} {\bf 96} (2017), no.~6 065022,
  [\href{http://arxiv.org/abs/1705.01027}{{\tt arXiv:1705.01027}}].

\bibitem{Pasterski:2017ylz}
S.~Pasterski, S.-H. Shao, and A.~Strominger, {\it {Gluon Amplitudes as 2d
  Conformal Correlators}},  {\em Phys. Rev. D} {\bf 96} (2017), no.~8 085006,
  [\href{http://arxiv.org/abs/1706.03917}{{\tt arXiv:1706.03917}}].

\bibitem{Donnay:2018neh}
L.~Donnay, A.~Puhm, and A.~Strominger, {\it {Conformally Soft Photons and
  Gravitons}},  {\em JHEP} {\bf 01} (2019) 184,
  [\href{http://arxiv.org/abs/1810.05219}{{\tt arXiv:1810.05219}}].

\bibitem{Fan:2019emx}
W.~Fan, A.~Fotopoulos, and T.~R. Taylor, {\it {Soft Limits of Yang-Mills
  Amplitudes and Conformal Correlators}},  {\em JHEP} {\bf 05} (2019) 121,
  [\href{http://arxiv.org/abs/1903.01676}{{\tt arXiv:1903.01676}}].

\bibitem{Pate:2019mfs}
M.~Pate, A.-M. Raclariu, and A.~Strominger, {\it {Conformally Soft Theorem in
  Gauge Theory}},  {\em Phys. Rev.} {\bf D100} (2019), no.~8 085017,
  [\href{http://arxiv.org/abs/1904.10831}{{\tt arXiv:1904.10831}}].

\bibitem{Adamo:2019ipt}
T.~Adamo, L.~Mason, and A.~Sharma, {\it {Celestial amplitudes and conformal
  soft theorems}},  {\em Class. Quant. Grav.} {\bf 36} (2019), no.~20 205018,
  [\href{http://arxiv.org/abs/1905.09224}{{\tt arXiv:1905.09224}}].

\bibitem{Puhm:2019zbl}
A.~Puhm, {\it {Conformally Soft Theorem in Gravity}},
  \href{http://arxiv.org/abs/1905.09799}{{\tt arXiv:1905.09799}}.

\bibitem{Guevara:2019ypd}
A.~Guevara, {\it {Notes on Conformal Soft Theorems and Recursion Relations in
  Gravity}},  \href{http://arxiv.org/abs/1906.07810}{{\tt arXiv:1906.07810}}.

\bibitem{Law:2019glh}
Y.~A. Law and M.~Zlotnikov, {\it {Poincar\'e constraints on celestial
  amplitudes}},  {\em JHEP} {\bf 20} (2020) 085,
  [\href{http://arxiv.org/abs/1910.04356}{{\tt arXiv:1910.04356}}].

\bibitem{Fotopoulos:2019vac}
A.~Fotopoulos, S.~Stieberger, T.~R. Taylor, and B.~Zhu, {\it {Extended BMS
  Algebra of Celestial CFT}},  {\em JHEP} {\bf 03} (2020) 130,
  [\href{http://arxiv.org/abs/1912.10973}{{\tt arXiv:1912.10973}}].

\bibitem{Banerjee:2020kaa}
S.~Banerjee, S.~Ghosh, and R.~Gonzo, {\it {BMS symmetry of celestial OPE}},
  {\em JHEP} {\bf 04} (2020) 130, [\href{http://arxiv.org/abs/2002.00975}{{\tt
  arXiv:2002.00975}}].

\bibitem{Fan:2020xjj}
W.~Fan, A.~Fotopoulos, S.~Stieberger, and T.~R. Taylor, {\it {On Sugawara
  construction on Celestial Sphere}},
  \href{http://arxiv.org/abs/2005.10666}{{\tt arXiv:2005.10666}}.

\bibitem{Nandan:2019jas}
D.~Nandan, A.~Schreiber, A.~Volovich, and M.~Zlotnikov, {\it {Celestial
  Amplitudes: Conformal Partial Waves and Soft Limits}},  {\em JHEP} {\bf 10}
  (2019) 018, [\href{http://arxiv.org/abs/1904.10940}{{\tt arXiv:1904.10940}}].

\bibitem{Gonzalez:2020tpi}
H.~A. Gonz\'alez, A.~Puhm, and F.~Rojas, {\it {Loops on the Celestial Sphere}},
   \href{http://arxiv.org/abs/2009.07290}{{\tt arXiv:2009.07290}}.

\bibitem{Casali:2020vuy}
E.~Casali and A.~Puhm, {\it {A Double Copy for Celestial Amplitudes}},
  \href{http://arxiv.org/abs/2007.15027}{{\tt arXiv:2007.15027}}.

\bibitem{Schreiber:2017jsr}
A.~Schreiber, A.~Volovich, and M.~Zlotnikov, {\it {Tree-level gluon amplitudes
  on the celestial sphere}},  {\em Phys. Lett. B} {\bf 781} (2018) 349--357,
  [\href{http://arxiv.org/abs/1711.08435}{{\tt arXiv:1711.08435}}].

\bibitem{Albayrak:2020saa}
S.~Albayrak, C.~Chowdhury, and S.~Kharel, {\it {On loop celestial amplitudes
  for gauge theory and gravity}},  \href{http://arxiv.org/abs/2007.09338}{{\tt
  arXiv:2007.09338}}.

\bibitem{Stieberger:2018edy}
S.~Stieberger and T.~R. Taylor, {\it {Strings on Celestial Sphere}},  {\em
  Nucl. Phys. B} {\bf 935} (2018) 388--411,
  [\href{http://arxiv.org/abs/1806.05688}{{\tt arXiv:1806.05688}}].

\bibitem{Banerjee:2020zlg}
S.~Banerjee, S.~Ghosh, and P.~Paul, {\it {MHV Graviton Scattering Amplitudes
  and Current Algebra on the Celestial Sphere}},
  \href{http://arxiv.org/abs/2008.04330}{{\tt arXiv:2008.04330}}.

\bibitem{Banerjee:2020vnt}
S.~Banerjee and S.~Ghosh, {\it {MHV Gluon Scattering Amplitudes from Celestial
  Current Algebras}},  \href{http://arxiv.org/abs/2011.00017}{{\tt
  arXiv:2011.00017}}.

\bibitem{Pate:2019lpp}
M.~Pate, A.-M. Raclariu, A.~Strominger, and E.~Y. Yuan, {\it {Celestial
  Operator Products of Gluons and Gravitons}},
  \href{http://arxiv.org/abs/1910.07424}{{\tt arXiv:1910.07424}}.

\bibitem{Fotopoulos:2019tpe}
A.~Fotopoulos and T.~R. Taylor, {\it {Primary Fields in Celestial CFT}},  {\em
  JHEP} {\bf 10} (2019) 167, [\href{http://arxiv.org/abs/1906.10149}{{\tt
  arXiv:1906.10149}}].

\bibitem{Ebert:2020nqf}
S.~Ebert, A.~Sharma, and D.~Wang, {\it {Descendants in celestial CFT and
  emergent multi-collinear factorization}},
  \href{http://arxiv.org/abs/2009.07881}{{\tt arXiv:2009.07881}}.

\bibitem{Mason:2013sva}
L.~Mason and D.~Skinner, {\it {Ambitwistor strings and the scattering
  equations}},  {\em JHEP} {\bf 07} (2014) 048,
  [\href{http://arxiv.org/abs/1311.2564}{{\tt arXiv:1311.2564}}].

\bibitem{Cachazo:2013hca}
F.~Cachazo, S.~He, and E.~Y. Yuan, {\it {Scattering of Massless Particles in
  Arbitrary Dimensions}},  {\em Phys. Rev. Lett.} {\bf 113} (2014), no.~17
  171601, [\href{http://arxiv.org/abs/1307.2199}{{\tt arXiv:1307.2199}}].

\bibitem{Cachazo:2013iea}
F.~Cachazo, S.~He, and E.~Y. Yuan, {\it {Scattering of Massless Particles:
  Scalars, Gluons and Gravitons}},  {\em JHEP} {\bf 07} (2014) 033,
  [\href{http://arxiv.org/abs/1309.0885}{{\tt arXiv:1309.0885}}].

\bibitem{Witten:2003nn}
E.~Witten, {\it {Perturbative gauge theory as a string theory in twistor
  space}},  {\em Commun. Math. Phys.} {\bf 252} (2004) 189--258,
  [\href{http://arxiv.org/abs/hep-th/0312171}{{\tt hep-th/0312171}}].

\bibitem{Berkovits:2004jj}
N.~Berkovits and E.~Witten, {\it {Conformal supergravity in twistor-string
  theory}},  {\em JHEP} {\bf 08} (2004) 009,
  [\href{http://arxiv.org/abs/hep-th/0406051}{{\tt hep-th/0406051}}].

\bibitem{Bern:2010ue}
Z.~Bern, J.~J.~M. Carrasco, and H.~Johansson, {\it {Perturbative Quantum
  Gravity as a Double Copy of Gauge Theory}},  {\em Phys. Rev. Lett.} {\bf 105}
  (2010) 061602, [\href{http://arxiv.org/abs/1004.0476}{{\tt
  arXiv:1004.0476}}].

\bibitem{Bern:2008qj}
Z.~Bern, J.~Carrasco, and H.~Johansson, {\it {New Relations for Gauge-Theory
  Amplitudes}},  {\em Phys. Rev. D} {\bf 78} (2008) 085011,
  [\href{http://arxiv.org/abs/0805.3993}{{\tt arXiv:0805.3993}}].

\bibitem{Mizera:2019blq}
S.~Mizera, {\it {Kinematic Jacobi Identity is a Residue Theorem: Geometry of
  Color-Kinematics Duality for Gauge and Gravity Amplitudes}},  {\em Phys. Rev.
  Lett.} {\bf 124} (2020), no.~14 141601,
  [\href{http://arxiv.org/abs/1912.03397}{{\tt arXiv:1912.03397}}].

\bibitem{Adamo:2017nia}
T.~Adamo, E.~Casali, L.~Mason, and S.~Nekovar, {\it {Scattering on plane waves
  and the double copy}},  {\em Class. Quant. Grav.} {\bf 35} (2018), no.~1
  015004, [\href{http://arxiv.org/abs/1706.08925}{{\tt arXiv:1706.08925}}].

\bibitem{Adamo:2017sze}
T.~Adamo, E.~Casali, L.~Mason, and S.~Nekovar, {\it {Amplitudes on plane waves
  from ambitwistor strings}},  {\em JHEP} {\bf 11} (2017) 160,
  [\href{http://arxiv.org/abs/1708.09249}{{\tt arXiv:1708.09249}}].

\bibitem{Adamo:2018mpq}
T.~Adamo, E.~Casali, L.~Mason, and S.~Nekovar, {\it {Plane wave backgrounds and
  colour-kinematics duality}},  {\em JHEP} {\bf 02} (2019) 198,
  [\href{http://arxiv.org/abs/1810.05115}{{\tt arXiv:1810.05115}}].

\bibitem{Adamo:2020syc}
T.~Adamo, L.~Mason, and A.~Sharma, {\it {MHV scattering of gluons and gravitons
  in chiral strong fields}},  {\em Phys. Rev. Lett.} {\bf 125} (2020), no.~4
  041602, [\href{http://arxiv.org/abs/2003.13501}{{\tt arXiv:2003.13501}}].

\bibitem{Adamo:2020qru}
T.~Adamo and A.~Ilderton, {\it {Classical and quantum double copy of
  back-reaction}},  {\em JHEP} {\bf 09} (2020) 200,
  [\href{http://arxiv.org/abs/2005.05807}{{\tt arXiv:2005.05807}}].

\bibitem{Adamo:2020yzi}
T.~Adamo, L.~Mason, and A.~Sharma, {\it {Gluon scattering on self-dual
  radiative gauge fields}},  \href{http://arxiv.org/abs/2010.14996}{{\tt
  arXiv:2010.14996}}.

\bibitem{Eberhardt:2020ewh}
L.~Eberhardt, S.~Komatsu, and S.~Mizera, {\it {Scattering Equations in AdS:
  Scalar Correlators in Arbitrary Dimensions}},
  \href{http://arxiv.org/abs/2007.06574}{{\tt arXiv:2007.06574}}.

\bibitem{Roehrig:2020kck}
K.~Roehrig and D.~Skinner, {\it {Ambitwistor Strings and the Scattering
  Equations on AdS$_3\times$S$^3$}},
  \href{http://arxiv.org/abs/2007.07234}{{\tt arXiv:2007.07234}}.

\bibitem{Law:2020tsg}
Y.~A. Law and M.~Zlotnikov, {\it {Massive Spinning Bosons on the Celestial
  Sphere}},  {\em JHEP} {\bf 06} (2020) 079,
  [\href{http://arxiv.org/abs/2004.04309}{{\tt arXiv:2004.04309}}].

\bibitem{Narayanan:2020amh}
S.~A. Narayanan, {\it {Massive Celestial Fermions}},
  \href{http://arxiv.org/abs/2009.03883}{{\tt arXiv:2009.03883}}.

\bibitem{Cachazo:2014xea}
F.~Cachazo, S.~He, and E.~Y. Yuan, {\it {Scattering Equations and Matrices:
  From Einstein To Yang-Mills, DBI and NLSM}},  {\em JHEP} {\bf 07} (2015) 149,
  [\href{http://arxiv.org/abs/1412.3479}{{\tt arXiv:1412.3479}}].

\bibitem{Cachazo:2013gna}
F.~Cachazo, S.~He, and E.~Y. Yuan, {\it {Scattering equations and
  Kawai-Lewellen-Tye orthogonality}},  {\em Phys. Rev. D} {\bf 90} (2014),
  no.~6 065001, [\href{http://arxiv.org/abs/1306.6575}{{\tt arXiv:1306.6575}}].

\bibitem{Cachazo:2014nsa}
F.~Cachazo, S.~He, and E.~Y. Yuan, {\it {Einstein-Yang-Mills Scattering
  Amplitudes From Scattering Equations}},  {\em JHEP} {\bf 01} (2015) 121,
  [\href{http://arxiv.org/abs/1409.8256}{{\tt arXiv:1409.8256}}].

\bibitem{Casali:2015vta}
E.~Casali, Y.~Geyer, L.~Mason, R.~Monteiro, and K.~A. Roehrig, {\it {New
  Ambitwistor String Theories}},  {\em JHEP} {\bf 11} (2015) 038,
  [\href{http://arxiv.org/abs/1506.08771}{{\tt arXiv:1506.08771}}].

\bibitem{Adamo:2013tsa}
T.~Adamo, E.~Casali, and D.~Skinner, {\it {Ambitwistor strings and the
  scattering equations at one loop}},  {\em JHEP} {\bf 04} (2014) 104,
  [\href{http://arxiv.org/abs/1312.3828}{{\tt arXiv:1312.3828}}].

\bibitem{Geyer:2015bja}
Y.~Geyer, L.~Mason, R.~Monteiro, and P.~Tourkine, {\it {Loop Integrands for
  Scattering Amplitudes from the Riemann Sphere}},  {\em Phys. Rev. Lett.} {\bf
  115} (2015), no.~12 121603, [\href{http://arxiv.org/abs/1507.00321}{{\tt
  arXiv:1507.00321}}].

\bibitem{Adamo:2015hoa}
T.~Adamo and E.~Casali, {\it {Scattering equations, supergravity integrands,
  and pure spinors}},  {\em JHEP} {\bf 05} (2015) 120,
  [\href{http://arxiv.org/abs/1502.06826}{{\tt arXiv:1502.06826}}].

\bibitem{Geyer:2015jch}
Y.~Geyer, L.~Mason, R.~Monteiro, and P.~Tourkine, {\it {One-loop amplitudes on
  the Riemann sphere}},  {\em JHEP} {\bf 03} (2016) 114,
  [\href{http://arxiv.org/abs/1511.06315}{{\tt arXiv:1511.06315}}].

\bibitem{Geyer:2016wjx}
Y.~Geyer, L.~Mason, R.~Monteiro, and P.~Tourkine, {\it {Two-Loop Scattering
  Amplitudes from the Riemann Sphere}},  {\em Phys. Rev. D} {\bf 94} (2016),
  no.~12 125029, [\href{http://arxiv.org/abs/1607.08887}{{\tt
  arXiv:1607.08887}}].

\bibitem{Geyer:2018xwu}
Y.~Geyer and R.~Monteiro, {\it {Two-Loop Scattering Amplitudes from Ambitwistor
  Strings: from Genus Two to the Nodal Riemann Sphere}},  {\em JHEP} {\bf 11}
  (2018) 008, [\href{http://arxiv.org/abs/1805.05344}{{\tt arXiv:1805.05344}}].

\bibitem{Adamo:2014wea}
T.~Adamo, E.~Casali, and D.~Skinner, {\it {A Worldsheet Theory for
  Supergravity}},  {\em JHEP} {\bf 02} (2015) 116,
  [\href{http://arxiv.org/abs/1409.5656}{{\tt arXiv:1409.5656}}].

\bibitem{Adamo:2018hzd}
T.~Adamo, E.~Casali, and S.~Nekovar, {\it {Yang-Mills theory from the
  worldsheet}},  {\em Phys. Rev. D} {\bf 98} (2018), no.~8 086022,
  [\href{http://arxiv.org/abs/1807.09171}{{\tt arXiv:1807.09171}}].

\bibitem{Adamo:2018ege}
T.~Adamo, E.~Casali, and S.~Nekovar, {\it {Ambitwistor string vertex operators
  on curved backgrounds}},  {\em JHEP} {\bf 01} (2019) 213,
  [\href{http://arxiv.org/abs/1809.04489}{{\tt arXiv:1809.04489}}].

\bibitem{Azevedo:2016zod}
T.~Azevedo and R.~L. Jusinskas, {\it {Background constraints in the infinite
  tension limit of the heterotic string}},  {\em JHEP} {\bf 08} (2016) 133,
  [\href{http://arxiv.org/abs/1607.06805}{{\tt arXiv:1607.06805}}].

\bibitem{Chandia:2015sfa}
O.~Chandia and B.~C. Vallilo, {\it {Ambitwistor pure spinor string in a type II
  supergravity background}},  {\em JHEP} {\bf 06} (2015) 206,
  [\href{http://arxiv.org/abs/1505.05122}{{\tt arXiv:1505.05122}}].

\bibitem{Bern:2019prr}
Z.~Bern, J.~J. Carrasco, M.~Chiodaroli, H.~Johansson, and R.~Roiban, {\it {The
  Duality Between Color and Kinematics and its Applications}},
  \href{http://arxiv.org/abs/1909.01358}{{\tt arXiv:1909.01358}}.

\bibitem{Mizera:2019gea}
S.~Mizera, {\it {Aspects of Scattering Amplitudes and Moduli Space
  Localization}},  other thesis, 2019.

\bibitem{Mizera:2016jhj}
S.~Mizera, {\it {Inverse of the String Theory KLT Kernel}},  {\em JHEP} {\bf
  06} (2017) 084, [\href{http://arxiv.org/abs/1610.04230}{{\tt
  arXiv:1610.04230}}].

\bibitem{10.2307/1969099}
N.~E. Steenrod, {\it {Homology With Local Coefficients}},  {\em Annals of
  Mathematics} {\bf 44} (1943), no.~4 610--627.

\bibitem{aomoto2011theory}
K.~Aomoto and M.~Kita, {\em {Theory of Hypergeometric Functions}}.
\newblock Springer Monographs in Mathematics. Springer Japan, 2011.

\bibitem{aomoto1977structure}
K.~Aomoto, {\it {On the structure of integrals of power product of linear
  functions}},  {\em Sci. Papers College Gen. Ed. Univ. Tokyo} {\bf 27} (1977),
  no.~2 49--61.

\bibitem{yoshida2013hypergeometric}
M.~Yoshida, {\em Hypergeometric functions, my love: modular interpretations of
  configuration spaces}, vol.~32.
\newblock Springer Science \& Business Media, 2013.

\bibitem{Stieberger:2018onx}
S.~Stieberger and T.~R. Taylor, {\it {Symmetries of Celestial Amplitudes}},
  {\em Phys. Lett.} {\bf B793} (2019) 141--143,
  [\href{http://arxiv.org/abs/1812.01080}{{\tt arXiv:1812.01080}}].

\bibitem{Aomoto}
K.~Aomoto and M.~Kita, {\em {Theory of hypergeometric functions}}.
\newblock Springer-Verlag, Tokyo, 2011.

\bibitem{Abe:2015ucn}
Y.~Abe, {\it {A note on generalized hypergeometric functions, KZ solutions, and
  gluon amplitudes}},  {\em Nucl. Phys. B} {\bf 907} (2016) 107--153,
  [\href{http://arxiv.org/abs/1512.06476}{{\tt arXiv:1512.06476}}].

\bibitem{Donnay:2020guq}
L.~Donnay, S.~Pasterski, and A.~Puhm, {\it {Asymptotic Symmetries and Celestial
  CFT}},  {\em JHEP} {\bf 09} (2020) 176,
  [\href{http://arxiv.org/abs/2005.08990}{{\tt arXiv:2005.08990}}].

\bibitem{Gelfand}
I.~M.~Gel'fand, {\it {General theory of hypergeometric functions}},  {\em
  Dokl.Akad.Nauk SSSR} {\bf 288} (1986) 14--18.

\bibitem{GKZ}
I.~M.~Gel'fand, M.~M.~Kapranov, and A.~V.~Zelevinsky, {\it {Generalized Euler
  integrals and A-hypergeometric functions}},  {\em Adv.Math.} {\bf 84} (1990)
  255.

\bibitem{delaCruz:2019skx}
L.~de~la Cruz, {\it {Feynman integrals as A-hypergeometric functions}},  {\em
  JHEP} {\bf 12} (2019) 123, [\href{http://arxiv.org/abs/1907.00507}{{\tt
  arXiv:1907.00507}}].

\bibitem{Arkani-Hamed:2019mrd}
N.~Arkani-Hamed, S.~He, and T.~Lam, {\it {Stringy Canonical Forms}},
  \href{http://arxiv.org/abs/1912.08707}{{\tt arXiv:1912.08707}}.

\bibitem{Ohmori:2015sha}
K.~Ohmori, {\it {Worldsheet Geometries of Ambitwistor String}},  {\em JHEP}
  {\bf 06} (2015) 075, [\href{http://arxiv.org/abs/1504.02675}{{\tt
  arXiv:1504.02675}}].

\bibitem{Adamo:2014yya}
T.~Adamo, E.~Casali, and D.~Skinner, {\it {Perturbative gravity at null
  infinity}},  {\em Class. Quant. Grav.} {\bf 31} (2014), no.~22 225008,
  [\href{http://arxiv.org/abs/1405.5122}{{\tt arXiv:1405.5122}}].

\bibitem{Adamo:2015fwa}
T.~Adamo and E.~Casali, {\it {Perturbative gauge theory at null infinity}},
  {\em Phys. Rev. D} {\bf 91} (2015), no.~12 125022,
  [\href{http://arxiv.org/abs/1504.02304}{{\tt arXiv:1504.02304}}].

\bibitem{Banerjee:2017jeg}
N.~Banerjee, S.~Banerjee, S.~Atul~Bhatkar, and S.~Jain, {\it {Conformal
  Structure of Massless Scalar Amplitudes Beyond Tree level}},  {\em JHEP} {\bf
  04} (2018) 039, [\href{http://arxiv.org/abs/1711.06690}{{\tt
  arXiv:1711.06690}}].

\end{thebibliography}\endgroup

\end{document}